\documentclass[aps,prd,twocolumn,epsf,nofootinbib,superscriptaddress,showpacs]{revtex4}

\usepackage{graphicx,here}
\usepackage{amsmath,amssymb,latexsym}
\usepackage{bm}
\usepackage{url}

\def\lsim{\mathrel{\raise.3ex\hbox{$<$\kern-.75em\lower1ex\hbox{$\sim$}}}}
\def\gsim{\mathrel{\raise.3ex\hbox{$>$\kern-.75em\lower1ex\hbox{$\sim$}}}}

\def\lbldef#1#2{\expandafter\gdef\csname #1\endcsname {#2}}

\def\href#1#2{#2}

\newcommand{\bwide}{\begin{widetext}}
\newcommand{\ewide}{\end{widetext}}
\newcommand{\beq}[1]{\begin{equation} \label{(#1)}}
\newcommand{\eeq}{\end{equation}}
\newcommand{\ba}[1]{\begin{eqnarray} \label{(#1)}}
\newcommand{\ea}{\end{eqnarray}}

\begin{document}


\title{Dark Matter Subhalos In the Fermi First Source Catalog}

\author{Matthew R.~Buckley}
\affiliation{Department of Physics, California Institute of Technology, Pasadena, CA 91125}
\author{Dan Hooper}
\affiliation{Center for Particle Astrophysics, Fermi National Accelerator Laboratory, Batavia, IL 60510}
\affiliation{Department of Astronomy and Astrophysics, The University of Chicago, Chicago, IL  60637} 

\begin{abstract}

The Milky Way's dark matter halo is thought to contain large numbers of smaller subhalos. These objects can contain very high densities of dark matter, and produce potentially observable fluxes of gamma rays. In this article, we study the gamma ray sources in the Fermi Gamma Ray Space Telescope's recently published First Source Catalog, and attempt to determine whether this catalog might contain a population of dark matter subhalos. We find that, while approximately 20-60 of the catalog's unidentified sources could plausibly be dark matter subhalos, such a population cannot be clearly identified as such at this time. From the properties of the sources in the First Source Catalog, we derive limits on the dark matter's annihilation cross section that are comparably stringent to those derived from recent observations of dwarf spheroidal galaxies. 

\end{abstract}

\pacs{95.35.+d;07.85.-m;98.70.Rz; FERMILAB-PUB-10-074-A; CALT 68-2785}

\maketitle

\section{Introduction}

In the standard theory of hierarchical structure formation, cold dark matter particles gather to form small halos, which later merge to form ever more massive halos. As a result of this process, dark matter halos (such that that hosting the Milky Way) are expected to contain many subhalos, ranging from dwarf spheroidal galaxies ($M \gsim 10^7 M_{\odot}$) down to objects with masses as small as $10^{-3}-10^{-8} M_{\odot}$. The minimum mass of dark matter subhalos is determined by the temperature at which the dark matter particles became kinetically decoupled from the cosmic neutrino background, which in turn depends on the particle physics of the dark matter candidate.  

With the exception of dwarf spheroidal galaxies, efforts to directly observe dark matter subhalos have failed to detect any such objects. A promising method to search for nearby dark matter substructures is to use gamma ray telescopes, such as the Fermi Gamma Ray Space Telescope (FGST), to detect the products of dark matter annihilations. To such a telescope, relatively large ($\sim 10^3-10^7 M_{\odot}$) and nearby ($\sim 0.01-10$ kpc) subhalos could appear as as bright and point-like gamma ray sources, without counterparts in other wavelengths. For reasonable assumptions (based on the results of numerical simulations) regarding the number of subhalos in the Milky Way and the dark matter distribution within those subhalos, one is led to expect a handful of subhalos to be observable by the FGST. For example, a dark matter candidate in the form of a 50 GeV particle with a annihilation cross section of $\sigma v \sim 3 \times 10^{-26}$ cm$^3$/s is predicted to provide on the order of a few subhalos that would be observable at or above the $5\sigma$ level~\cite{Pieri:2007ir}. 

The FGST collaboration has recently published a catalog of 1451 point sources~\cite{catalog}, including 630 that are not associated with sources in other astronomical catalogs. Of these unidentified sources, 368 have been detected with greater than 5$\sigma$ significance, and are more than 10 degrees away from the Galactic Plane. In this article, we study this collection of gamma ray sources and consider whether any significant fraction might be the result of dark matter annihilations taking place in nearby subhalos. 

The remainder of this article is structured as follows. In Sec.~\ref{theory}, we calculate the number of dark matter subhalos predicted to be observable by FGST, for various astrophysical assumptions and particle dark matter candidates. In Sec.~\ref{catalog}, we determine which of the objects contained within the FGST's First Source Catalog can be fit by dark matter annihilations, for various combinations of mass and annihilation channel. We use this information in Sec.~\ref{limits} to derive upper limits on the dark matter's annihilation cross section. In Sec.~\ref{hints}, we discuss some of the more interesting features of the First Source Catalog, and consider whether they might result from a population of relatively large and nearby subhalos. We summarize our results in Sec.~\ref{conclusions}.

\section{Gamma Rays From Nearby Dark Matter Subhalos}
\label{theory}

Following the results of numerical simulations, we begin by assuming that the Galactic Halo contains dark matter subhalos with a mass function given by $dN_n/dM_h \propto M_h^{-2}$, down to some minimum mass, normalized such that 10\% of the total mass of the halo is found in $10^7$-$10^{10} \, M_{\odot}$ subhalos~\cite{norm}. For a minimum subhalo mass of one Earth mass, this normalization corresponds to a total of $\sim 5\times 10^{16}$ subhalos within the Milky Way, which collectively make up about half of our Galaxy's total mass. In the local neighborhood, this implies a number density of approximately 34 (roughly Earth mass) subhalos per cubic parsec. The precise value for the mimimum subhalo mass depends on the characteristics of the dark matter particle~\cite{smallest}. As this study relies primarily on the properties of the largest mass subhalos, the minimum mass adopted has no significant impact on our results.

We further assume that the subhalos are described by the following density profile, as supported by the Via Lactea simulation~\cite{1pt2}:
\begin{equation}
\rho(r) \propto \frac{1}{(r/R_s)^{1.2}\, [1+(r/R_s)]^2},
\label{1pt2}
\end{equation}
where $R_s$ is the scale radius of the subhalo. One should note, however, that the halo profiles describing the inner volumes of dark matter subhalos are difficult to resolve using present simulations, and groups other than Via Lactea collaboration (notably, the Aquarius Project) have reported somewhat less steep inner profiles~\cite{shallow}. With this in mind, we will also calculate results using an Einasto {\it et al.} profile~\cite{einasto}:
\begin{equation}
\rho(r) \propto \exp \bigg[ -\frac{2}{\alpha} \bigg(\frac{r}{R_s}\bigg)^{\alpha}\bigg],
\end{equation}
where $\alpha \approx 0.17$.

The concentration of a halo is defined as the ratio of its virial radius to its scale radius. To estimate the concentration of a halo of a given mass, we use the analytic model of Bullock {\it et al.}~\cite{bullock}. This model estimates a concentration of 21 (27) for a subhalo of mass $10^{7} M_{\odot}$ ($10^{5} M_{\odot}$). For large halo masses, the results of numerical simulations are in good agreement with this model. Smaller subhalos, currently beyond the resolution of such simulations, are largely irrelevant to our study. There are also indications from numerical simulations that subhalo in the inner volumes of their host halo tend to have, on average, higher concentrations~\cite{Kuhlen:2008aw}. To be conservative we do not include this effect in our calculations.

One should keep in mind that considerable halo-to-halo variation in the concentration and shape of subhalo profiles has been observed in numerical simulations. The model of Bullock {\it et al.}, for example, only provides a measure of the average concentration of a subhalo of a given mass. The probability of a halo having a given concentration can be modelled by a lognormal distribution, with a dispersion of $\sigma_c\approx 0.24$~\cite{bullock}. Although we adopt mean concentration values in our calculations, we estimate that halo-to-halo fluctuations will increase the number of observable subhalos by a factor of about two in most cases (see also Ref.~\cite{Pieri:2007ir}). Note that, throughout this study, references to subhalo masses denote the mass prior to loss through tidal stripping.

\begin{figure*}[t]
\begin{center}
{\includegraphics[angle=0,width=0.48\linewidth]{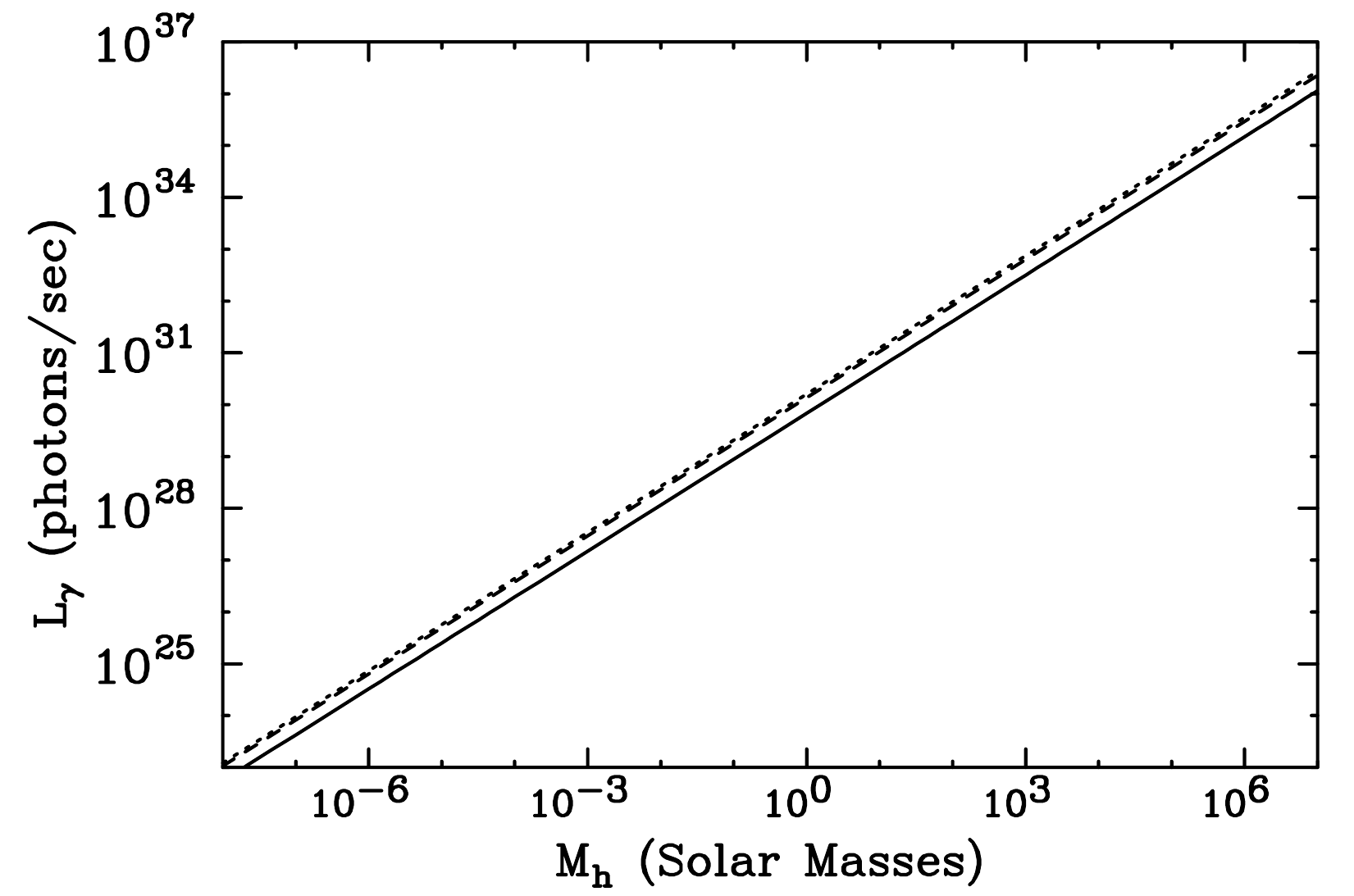}}
\hspace{0.02\linewidth}
{\includegraphics[angle=0,width=0.48\linewidth]{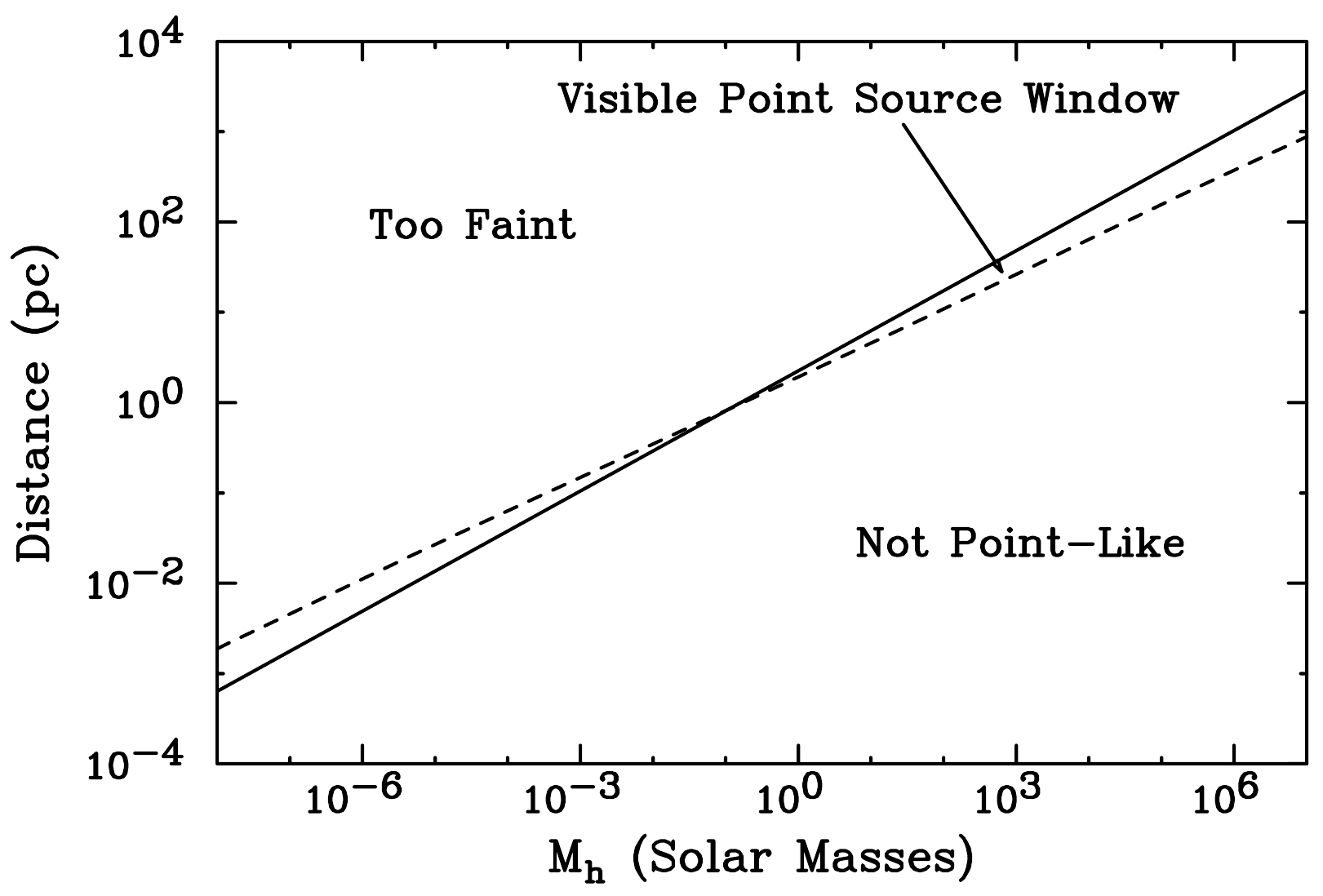}}
\vspace{-0.3cm}
\caption{Left frame: The number of gamma rays per second produced from dark matter annihilations in a subhalo as a function of the subhalo's mass (prior to any mass loss). The solid and dashed lines correspond to the cases in which the outermost 99\% or 90\% of the original mass of the subhalo has been lost through tidal interactions, respectively. The dotted line (which may be difficult to distinguish from the dashed line) represents the zero mass loss case. Right frame: The range of halo masses (again, prior to any mass loss) and distances for which the gamma ray annihilation products from a subhalo may be detectable by FGST and appear approximately as a point source. To be considered potentially observable, we require more than 50 events per year above 1 GeV from a given subhalo. We have assumed that 99\% of the original subhalo's mass has been lost. To be considered point-like, we require that more than 95\% of the photons from a subhalo be concentrated within $2^{\circ}$ (approximately the 95\% containment angle for 1 GeV photons at FGST). In each frame, we have adopted a dark matter profile as described in Eq.~\ref{1pt2}, and considered a dark matter particle with $m_{\rm DM}=50$ GeV, $\sigma v=3\times 10^{-26}$ cm$^3$/s (the typical value for a thermal relic), and that annihilates primarily to $b\bar{b}$.}
\label{window}
\end{center}
\end{figure*}

Subhalos in the local volume of the Milky Way are likely to have had a large fraction of their mass stripped through tidal interactions with other halos and stars. This primarily impacts a subhalo's outer density profile, leaving its denser, more tightly bound inner cusp intact~\cite{disruption}. As the default assumption throughout our calculations, we assume that nearby subhalos have lost the outermost 99\% of their total mass (although this only modestly impacts the overall annihilation rate). 

Just as the halo of the Milky Way contains many subhalos, subhalos themselves are also expected to contain smaller bound dark matter structures within their volumes. This can lead to an overall ``boost factor'' to the dark matter annihilation rate from such objects. To be conservative, we do not include any such boost in our (default) calculations.

The rate of gamma rays produced from dark matter annihilations taking place in a nearby subhalo is given by
\begin{equation}
L_{\gamma}=\frac{\sigma v}{2 m^2_{\rm DM}} \int \frac{dN_{\gamma}}{dE_{\gamma}} dE_{\gamma} \int \rho^2 dV,
\end{equation}
where $\sigma v$ and $m_{\rm DM}$ are the annihilation cross section and mass of the dark matter particle, respectively, and the second integral is performed over the volume of the subhalo. $dN_{\gamma}/dE_{\gamma}$ is the spectrum of gamma rays produced per dark matter annihilation, which depends on the dominant annihilation channel(s) and on the mass of the dark matter particle (we use PYTHIA~\cite{pythia} as implemented in DARKSUSY \cite{Gondolo:2004sc} to calculate the gamma ray spectrum). In the left frame of Fig.~\ref{window}, we show the gamma ray luminosity from a subhalo, as a function of the subhalo's mass. Results are shown for 99\%, 90\% and 0\% mass loss.

In order for the gamma ray annihilation products from a subhalo to constitute a source that could potentially appear in the FGST point source catalog, the subhalo must be both sufficiently bright and sufficiently compact to mimic a point source. To estimate the number of events from a subhalo observed by FGST, we multiply the gamma ray flux by an effective area of 6800 cm$^2$, and a coverage of 20\% of the sky at any given time. Although the detectability of a given gamma ray source depends somewhat on its spectral shape and its location in the sky, we can roughly estimate how bright a given subhalo must be to be detectable at high significance by FGST. In particular, the diffuse gamma ray flux measured by FGST generates approximately 20 events per year per square degree above 1 GeV over galactic latitudes of $10^{\circ}<|b|<20^{\circ}$, and about 60-70 events per year per square degree above 1 GeV over galactic latitudes of $|b| > 60^{\circ}$. In these two regions of sky, we estimate that $5\sqrt{20} \approx 20$ or $5\sqrt{60} \approx 40$ signal events per year above 1 GeV would be required in order for a subhalo to be potentially discovered with 5$\sigma$ significance. Based on this estimate, we conservatively classify any subhalo that produces more than 50 events above 1 GeV per year at FGST to be potentially detectable. In the right frame of Fig.~\ref{window}, subhalos below the solid line are sufficiently bright to be potentially detected by FGST according to this criteria. Although the smallest subhalos (sometimes called ``microhalos'') are much more numerous, individual sub-solar mass halos are unlikely to provide observable fluxes gamma rays (also see, for example, Ref.~\cite{subsolar}). We find that most of the subhalos potentially detectable by FGST have relatively large masses, $M \sim 10^5-10^7 M_{\odot}$.

To appear as a point source to the FGST, the angular extent of the halo must not be much larger than the telescope's point spread function. To be considered point-like, we require that 95\% of the photons from a subhalo come from within a 2$^{\circ}$ radius (approximately the 95\% containment angle for 1 GeV photons at FGST). In Fig.~\ref{window}, only those halos below the solid line and above the dashed line are both sufficiently bright and point-like to potentially appear in the FGST point source catalog. Here, we have considered a 50 GeV dark matter particle that annihilates to $b\bar{b}$ with a cross section of $\sigma v = 3 \times 10^{-26}$ cm$^3$/s.

\begin{figure*}[t]
\begin{center}
{\includegraphics[angle=0,width=0.7\linewidth]{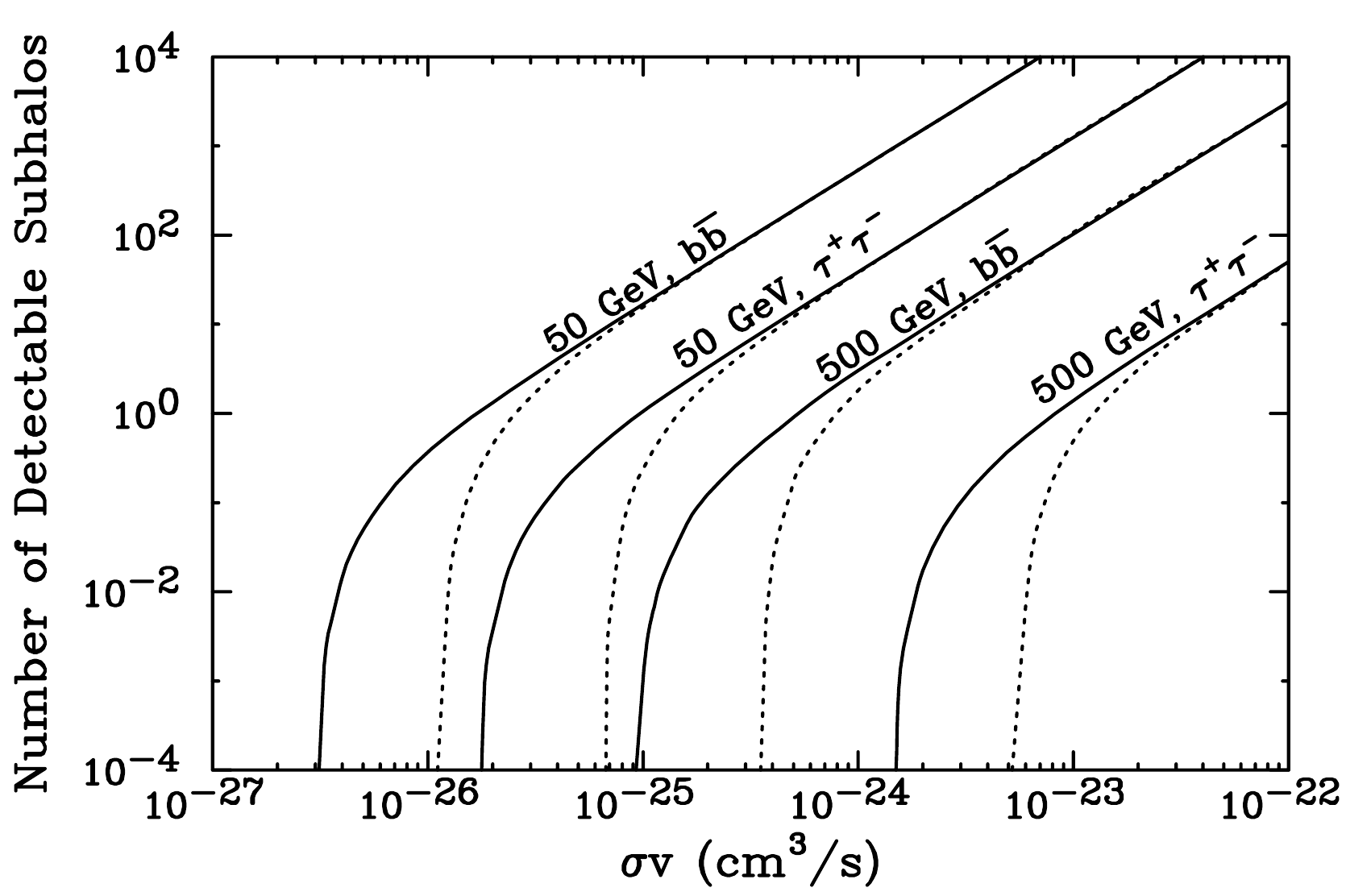}}
\vspace{-0.3cm}
\caption{The number of point-like subhalos potentially detectable by the Fermi Gamma Ray Space Telescope as a function of dark matter's annihilation cross section. To qualify as detectable and point-like, we require a subhalo to produce more than 50 events above 1 GeV per year at FGST, and require that more than 95\% of those photons be concentrated within a radius of $2^{\circ}$ (solid) or $1^{\circ}$ (dotted). We show results for two dark matter masses (50 and 500 GeV), and two annihilation channels ($b\bar{b}$ and $\tau^+ \tau^-$).}
\label{number}
\end{center}
\end{figure*}

For the case of a 50 GeV dark matter particle that annihilates to $b\bar{b}$ with $\sigma v = 3 \times 10^{-26}$ cm$^3$/s, we estimate that FGST is expected to detect 50 or more photons with energy $>$1 GeV from only a few point-like subhalos in a year of observation. This result, however, depends strongly on the mass, annihilation cross section, and annihilation channels of the dark matter particle. If we increase the cross section to $\sigma v = 10^{-25}$ cm$^3$/s (approximately the upper limit allowed by observations of dwarf spheroidal galaxies~\cite{fgstdwarf}), for example, we expect FGST to detect approximately 16 subhalos, most of which with masses between $10^6$ and $10^7$ solar masses, prior to mass losses (we consider only halos up to $10^7$ solar masses in our calculations). In Fig.~\ref{number}, we show the number of subhalos detectable by FGST as a function of annihilation cross section, for selected choices of the dark matter's mass and dominant annihilation channel.

At this point, we would like to comment on the uncertainties involved in estimating the number of subhalos that could potentially appear within FGST's point source catalog. The details of the subhalo profiles, concentrations, and the overall fraction of dark matter mass contained in subhalos can each significantly impact the number of subhalos that are observable by FGST. In Table~\ref{table}, we give the number of point-like subhalos observable by FGST for a variety of possible astrophysical assumptions, each for the case of a 50 GeV dark matter particle that annihilates to $b\bar{b}$ with $\sigma v = 10^{-25}$ cm$^3$/s. By varying the inner slope of the subhalos' density profile and the fraction of mass that is lost from the original subhalo (ML), we find that the number of observable subhalos can vary by a factor of several in either direction from the results shown in Fig.~\ref{number}.

If we consider the effects of substructure within subhalos themselves, we find that extrapolating the Bullock {\it et al.} concentration estimates down to a minimum mass of $10^{-6} M_{\odot}$ ($10^{-8} M_{\odot}$) can further enhance the number of observable subhalos by a factor of 2.4 (5.0). This range of estimates and uncertainties is consistent with those found, for example, in Ref.~\cite{Pieri:2007ir}. Furthermore, halo-to-halo variation in the concentration and shape of dark matter subhalos is expected to increase the number of observable subhalos by a factor of about two~\cite{Pieri:2007ir}. For this reason, we consider the results shown in Fig.~\ref{number} and Table~\ref{table} to be conservative.

\begin{table}[tbhp]
\centering
\begin{tabular}{|c|c|c|} \hline
Model/Parameters & Detectable, Point-Like Subhalos \\ \hline\hline
Default ($\gamma=1.2$, ML=99\%) & 16.8\\  \hline
$\gamma=1.0$, ML=99\% & 1.94\\  \hline
$\gamma=1.2$, ML=90\% & 46.4\\  \hline
$\gamma=1.0$, ML=90\% & 8.57\\  \hline
Einasto $\alpha=0.17$, ML=99\% & 6.50\\  \hline
Einasto $\alpha=0.17$, ML=90\% & 13.4\\  \hline
\hline
Default w/ sub-subhalos &  \\ \hline
$M_{\rm min}=10^{-6} M_{\odot}$ & 40.9 \\ \hline
$M_{\rm min}=10^{-8} M_{\odot}$ & 83.9 \\ \hline \hline
\end{tabular}
\caption{\label{table} The impact of various astrophysical assumptions on the number of point-like subhalos observable by FGST. Results are shown for a 50 GeV dark matter particle that annihilates with $\sigma v =  10^{-25}$ cm$^3$/s to $b \bar{b}$. Note that halo-to-halo variation in the concentration parameter is expected to further increase the number of detectable point-like halos by a factor of $\sim$2. See text for more details.}
\end{table}

\section{The Fermi Point Source Catalog}
\label{catalog}

In this section, we turn our attention to the FGST First Source Catalog~\cite{catalog}, which identifies point sources with greater than $\sim4\sigma$ significance within the data collected between August 2008 and July 2009, over the 100~MeV to 100 GeV energy range. Within the context of dark matter subhalos, we are interested in sources that are non-variable and are not associated with known astrophysical objects. Furthermore, we only consider point sources that are more than $10^\circ$ away from the Galactic Plane (that is, $|b|>10^{\circ}$), as it is expected that the region close to the plane will contain the majority of baryonic gamma ray sources (pulsars, supernova remnants, X-ray binaries, etc.). We also only consider sources that have been detected at greater than 5$\sigma$ significance. Once these criteria have been applied, there remain 368 objects in the FGST First Source Catalog that qualify as potential dark matter subhalo candidates (from a total of 1451 sources). Of these, 41 have been detected with greater than $10\sigma$ significance.

\begin{figure*}[t]
\begin{center}
{\includegraphics[width=0.6\linewidth]{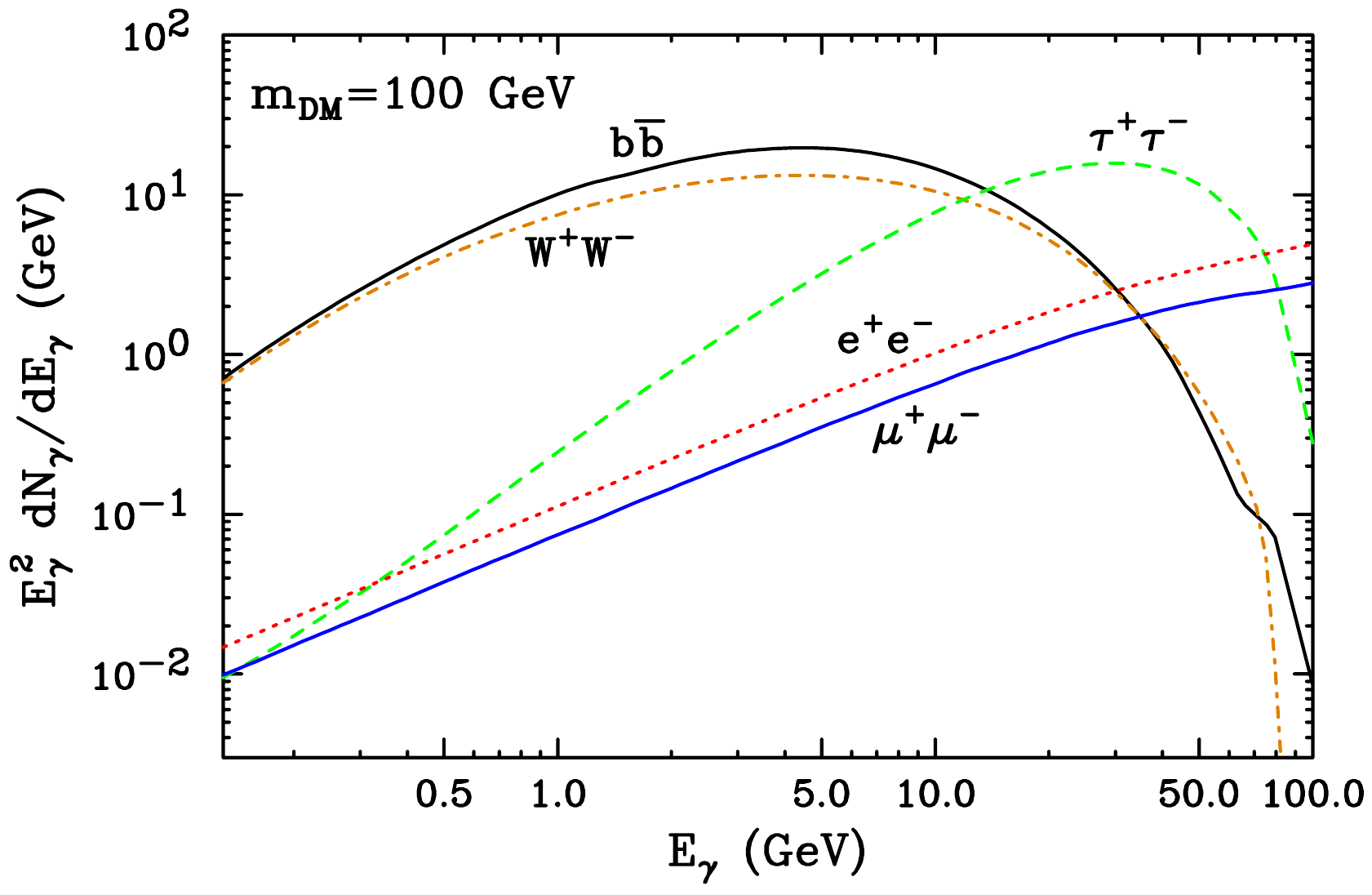}}
\vspace{-0.3cm}
\caption{The gamma ray spectrum per dark matter annihilation for $m_{\rm DM} =100$~GeV, and five dominant annihilation channels: $b\bar{b}$ (solid black), $W^+W^-$ (dot-dashed orange), $e^+e^-$ (dotted red), $\mu^+\mu^-$ (solid blue), and $\tau^+\tau^-$ (dashed yellow).\label{fig:spectrum}}
\end{center}
\end{figure*}

\begin{figure*}[t]
\begin{center}
{\includegraphics[angle=0,width=0.46\linewidth]{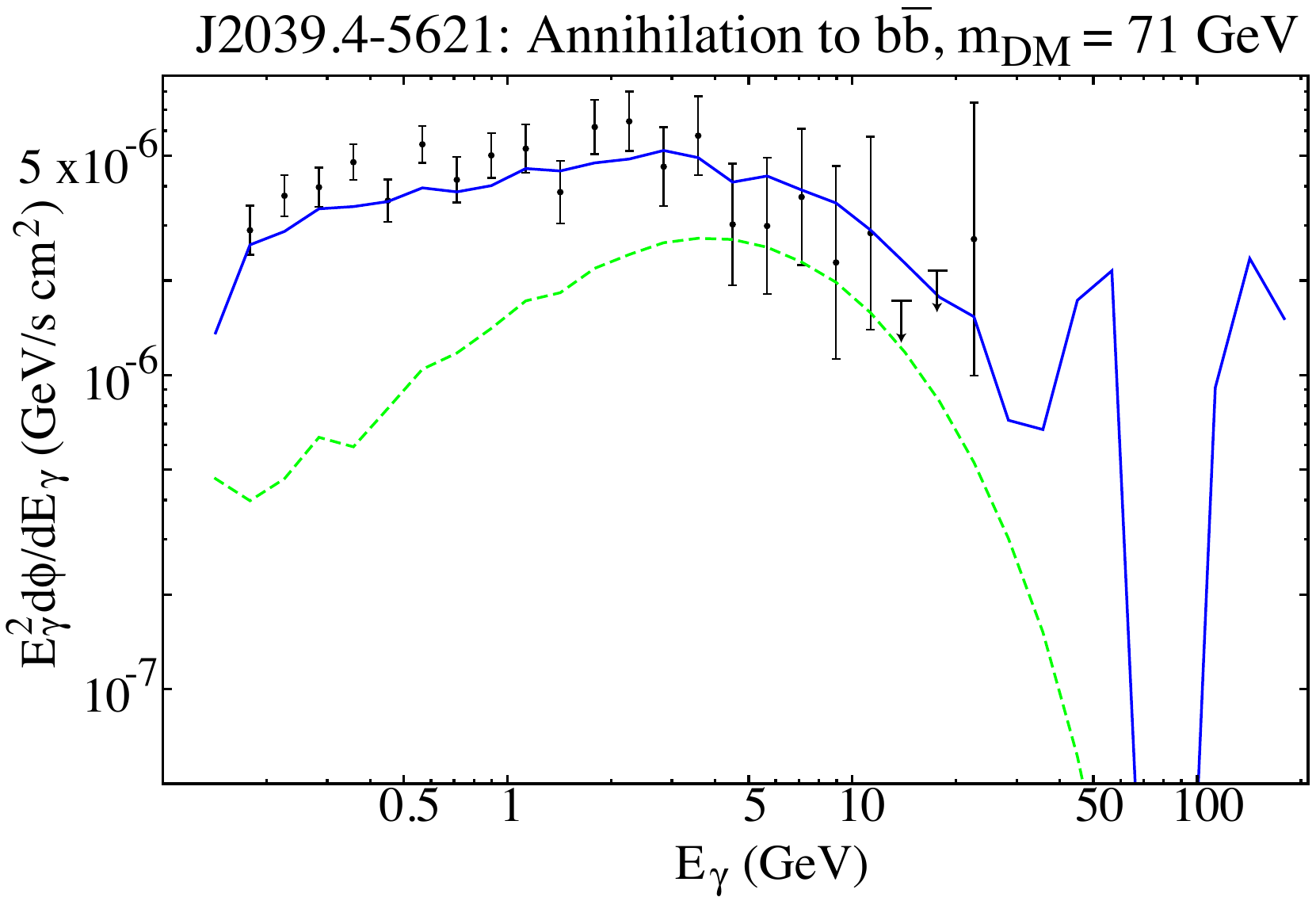}}
\hspace{0.02\linewidth}
{\includegraphics[angle=0,width=0.46\linewidth]{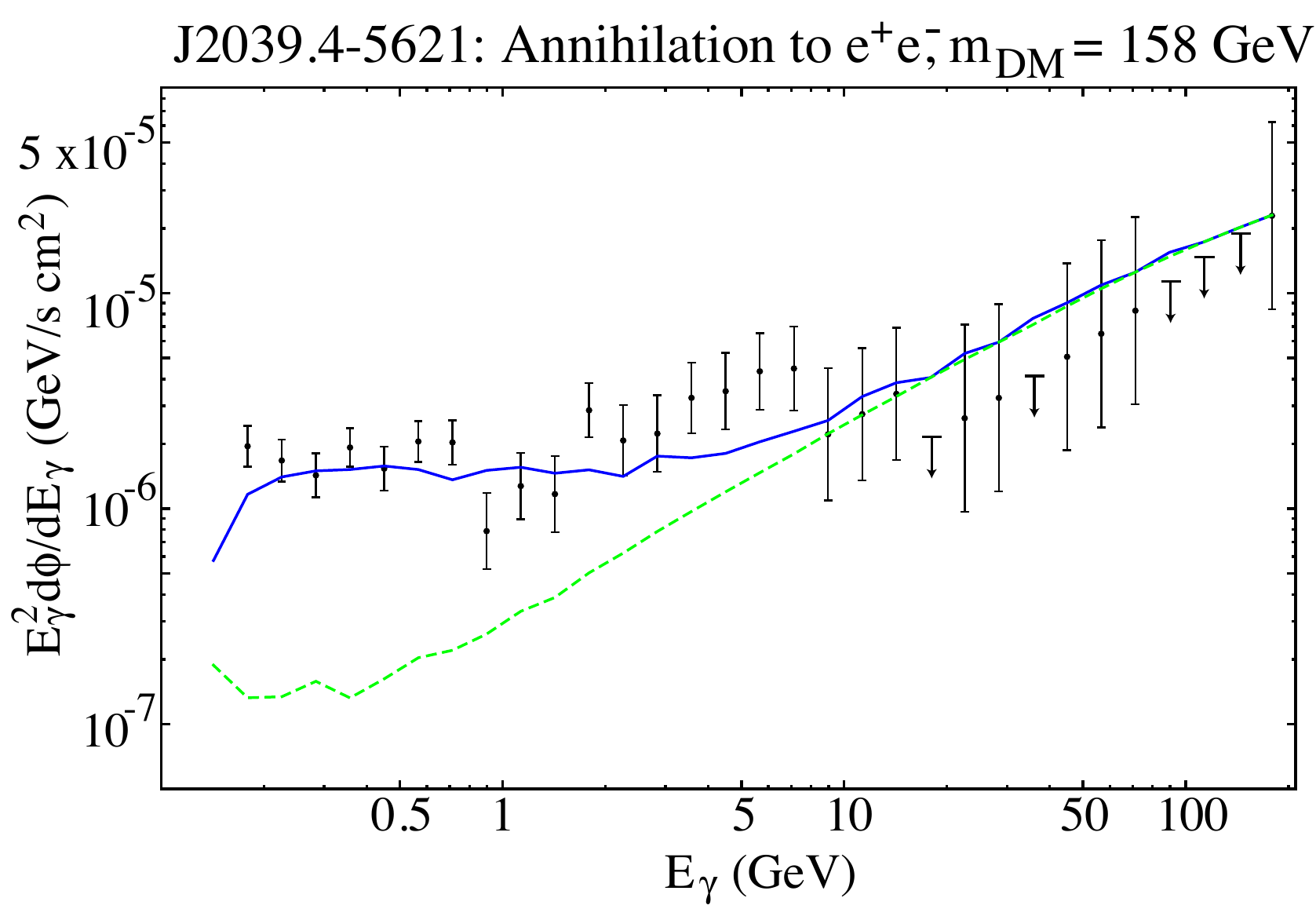}}\\
{\includegraphics[angle=0,width=0.46\linewidth]{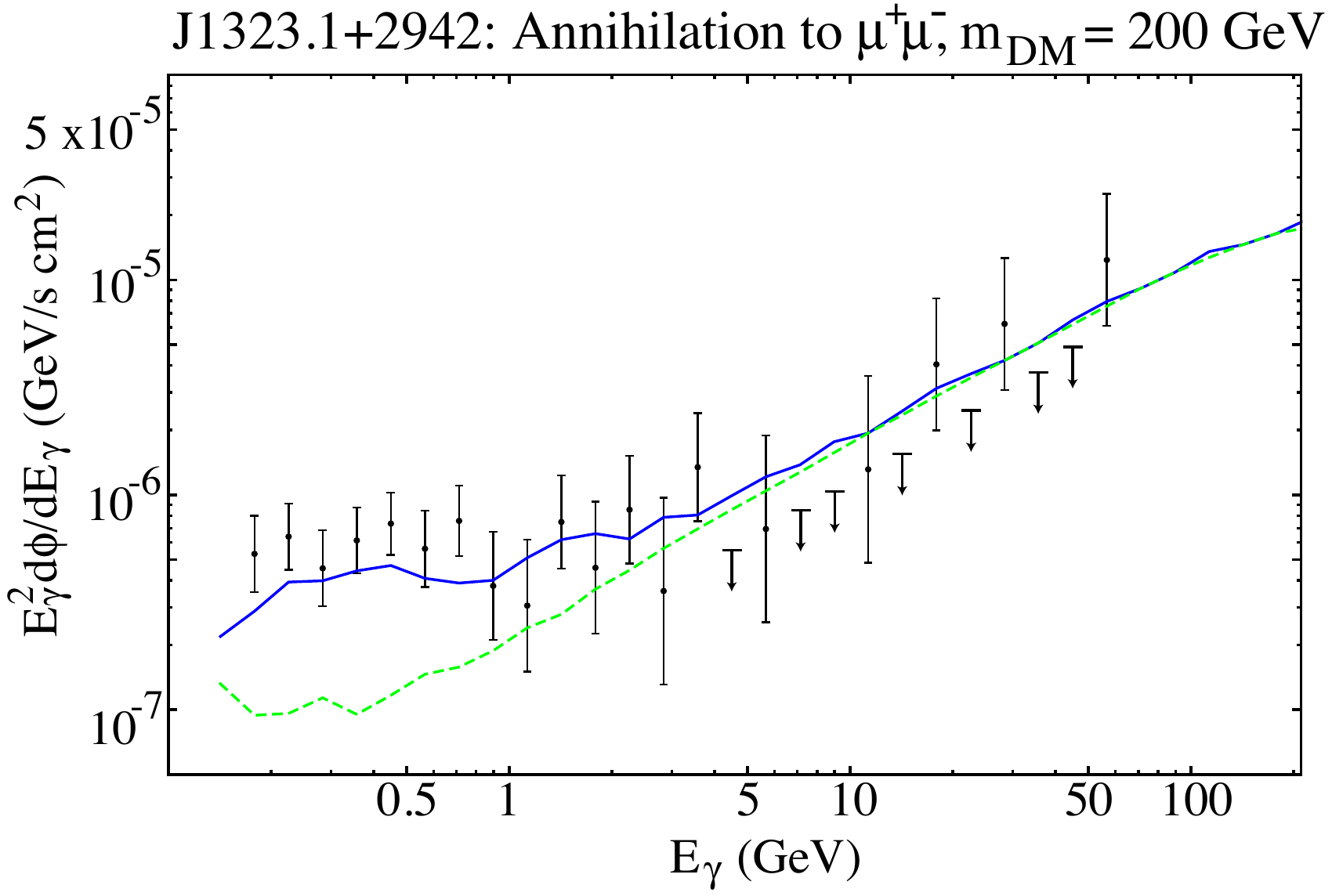}}
\hspace{0.02\linewidth}
{\includegraphics[angle=0,width=0.46\linewidth]{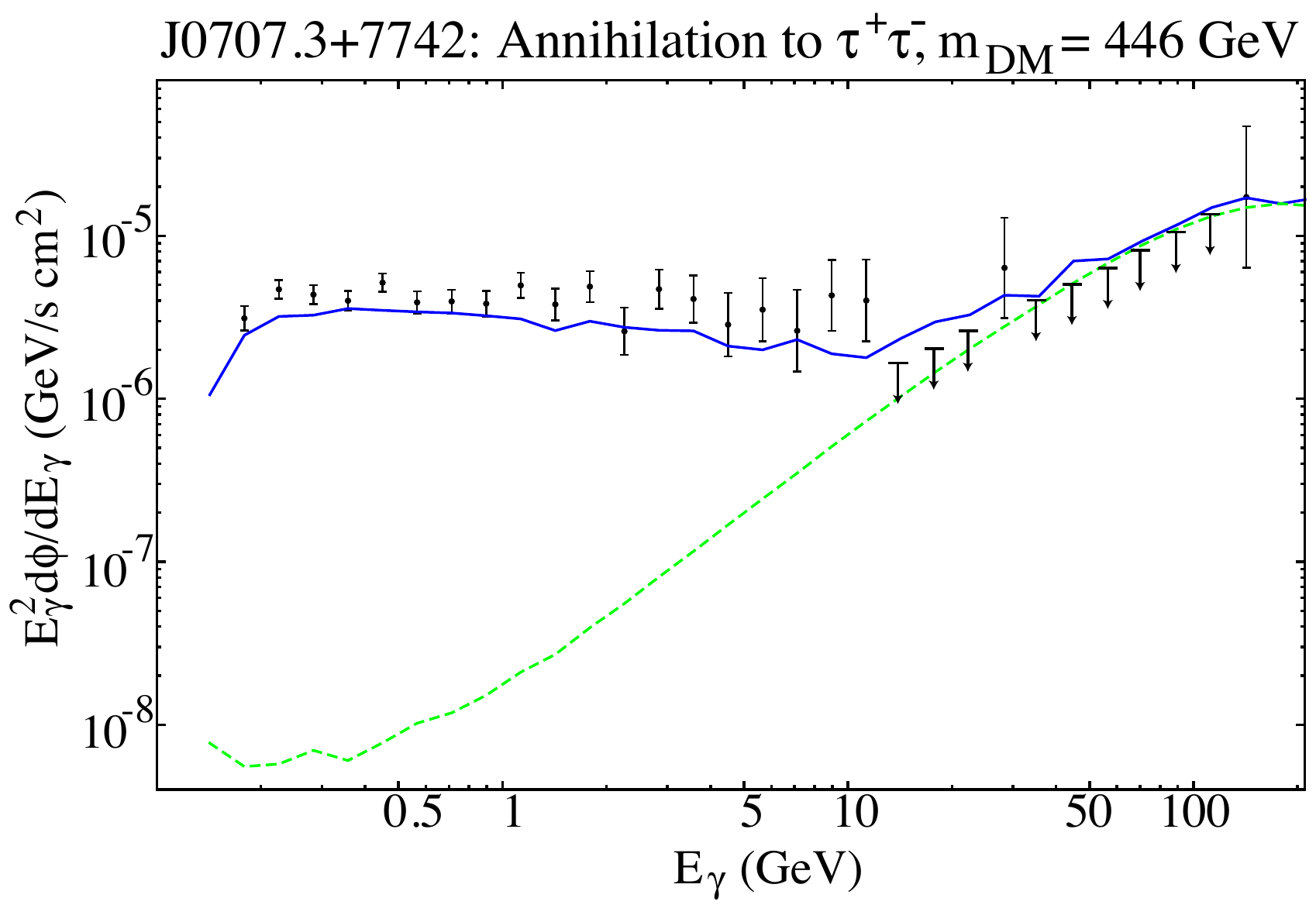}}\\
{\includegraphics[angle=0,width=0.46\linewidth]{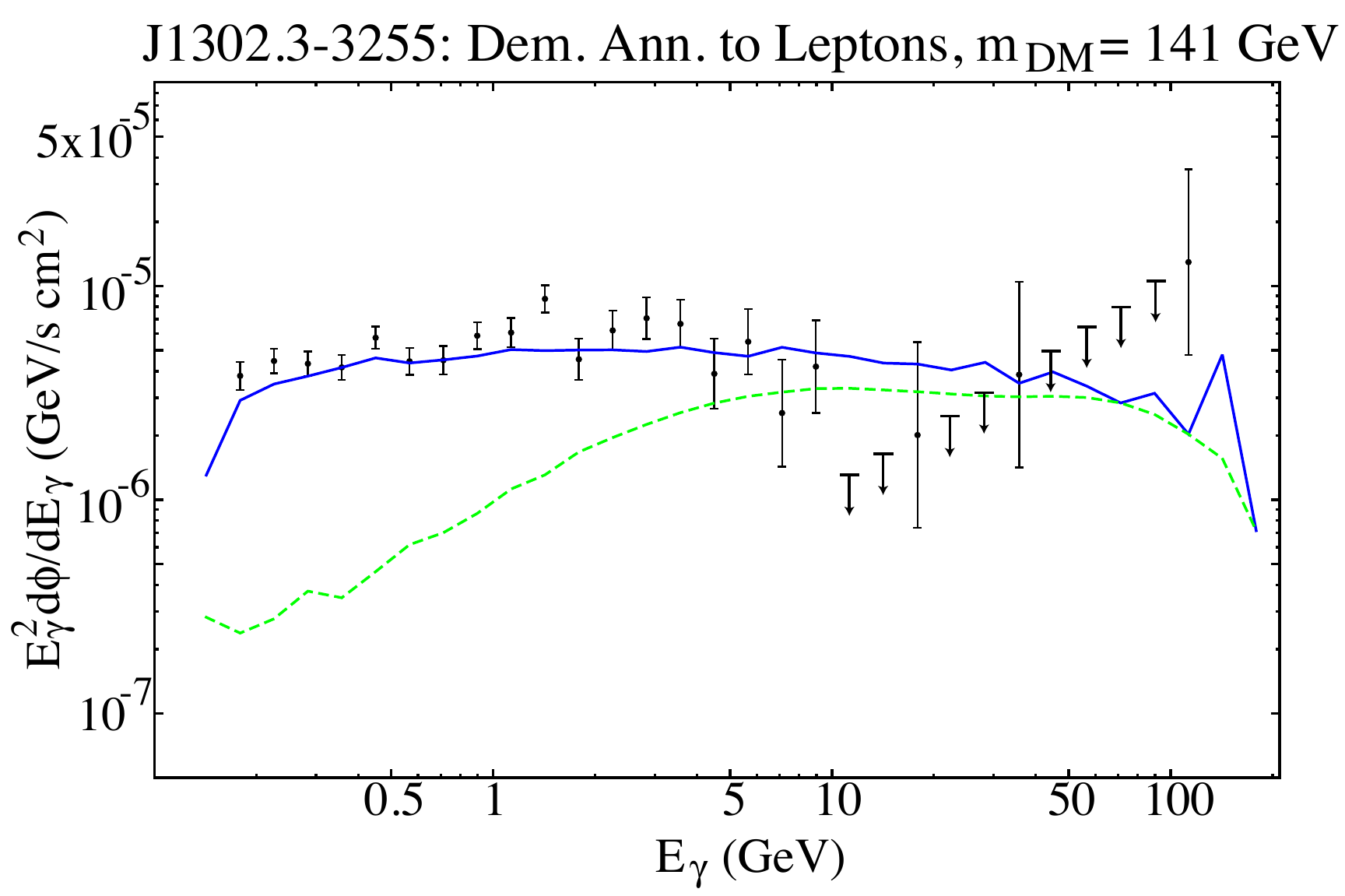}}
\hspace{0.02\linewidth}
{\includegraphics[angle=0,width=0.46\linewidth]{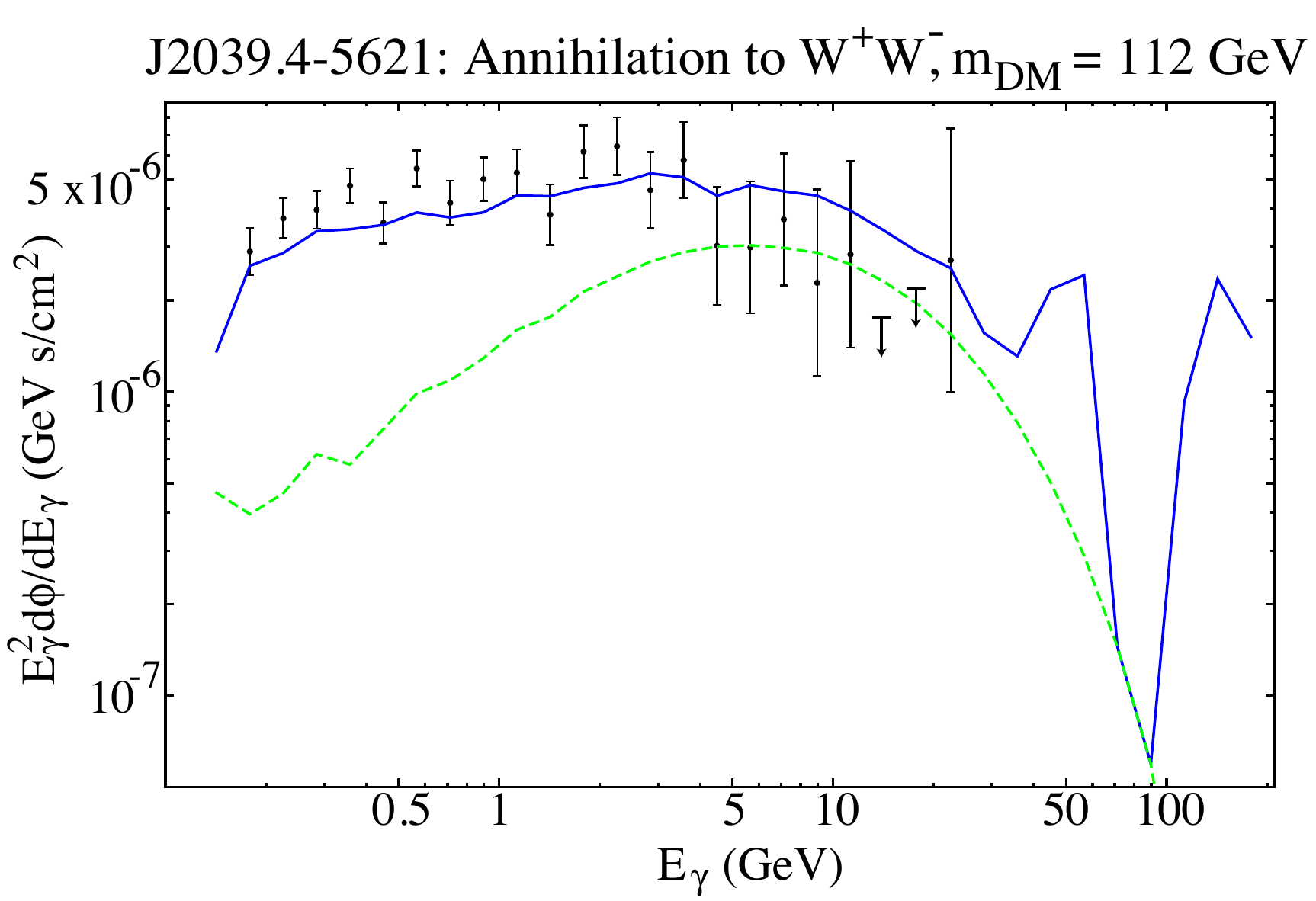}}
\caption{Examples of annihilating dark matter fit to the spectra of objects in the Fermi First Source Catalog. The black data points and error bars denote the spectrum observed by FGST, while the dotted green line is the spectrum from dark matter annihilations, and the blue line is the sum of the background and the annihilation spectrum. See text for more details.}
\label{fig:spectrumfits}
\end{center}
\end{figure*}

For each of the 368 subhalo candidate point sources, we compared their observed spectrum to the spectrum predicted from annihilating dark matter. To determine the observed spectrum from a given source, we took the gamma rays detected from a signal region within an angle equal to the 95\% point spread function from the identified center of the object, and subtracted the average spectrum over an annulus between $3^\circ$ and $5^\circ$ around the source. Throughout our analysis, the spectra were binned in 34 logarithmic bins distributed between 120~MeV and 251~GeV.  For many of the point sources, some of the bins, especially at high energies, contained no events. Note that, as the point spread function of FGST varies considerably with energy, the angular size of the signal region is smaller for higher energy bins~\cite{latper}. Events with a zenith angle greater than $105^\circ$ were excluded from our analysis, as this region is contaminated by gamma rays from the limb of the Earth~\cite{catalog}.

Using the publicly available program PYTHIA~\cite{pythia}, as implemented within DARKSUSY~\cite{Gondolo:2004sc}, we calculated the spectrum of gamma rays from dark matter annihilation for 60 different masses between 10~GeV and 10 TeV. Spectra were calculated for six annihilation channels: $b\bar{b}$, $e^+e^-$, $\mu^+\mu^-$, $\tau^+\tau^-$, `democratic leptons' (equal numbers of annihilations to $e^+e^-$, $\mu^+\mu^-$, and $\tau^+\tau^-$), and $W^+W^-$. For the gamma ray spectrum from dark matter annihilations to electrons, we calculated the final state radiation using the analytic formula described in Ref.~\cite{Bergstrom:2004cy}.
In Fig.~\ref{fig:spectrum}, we show the gamma ray spectrum per annihilation for the case of a 100 GeV dark matter particle, for several of the annihilation channels.  




\begin{figure*}[t]
\begin{center}
{\includegraphics[angle=0,width=0.46\linewidth]{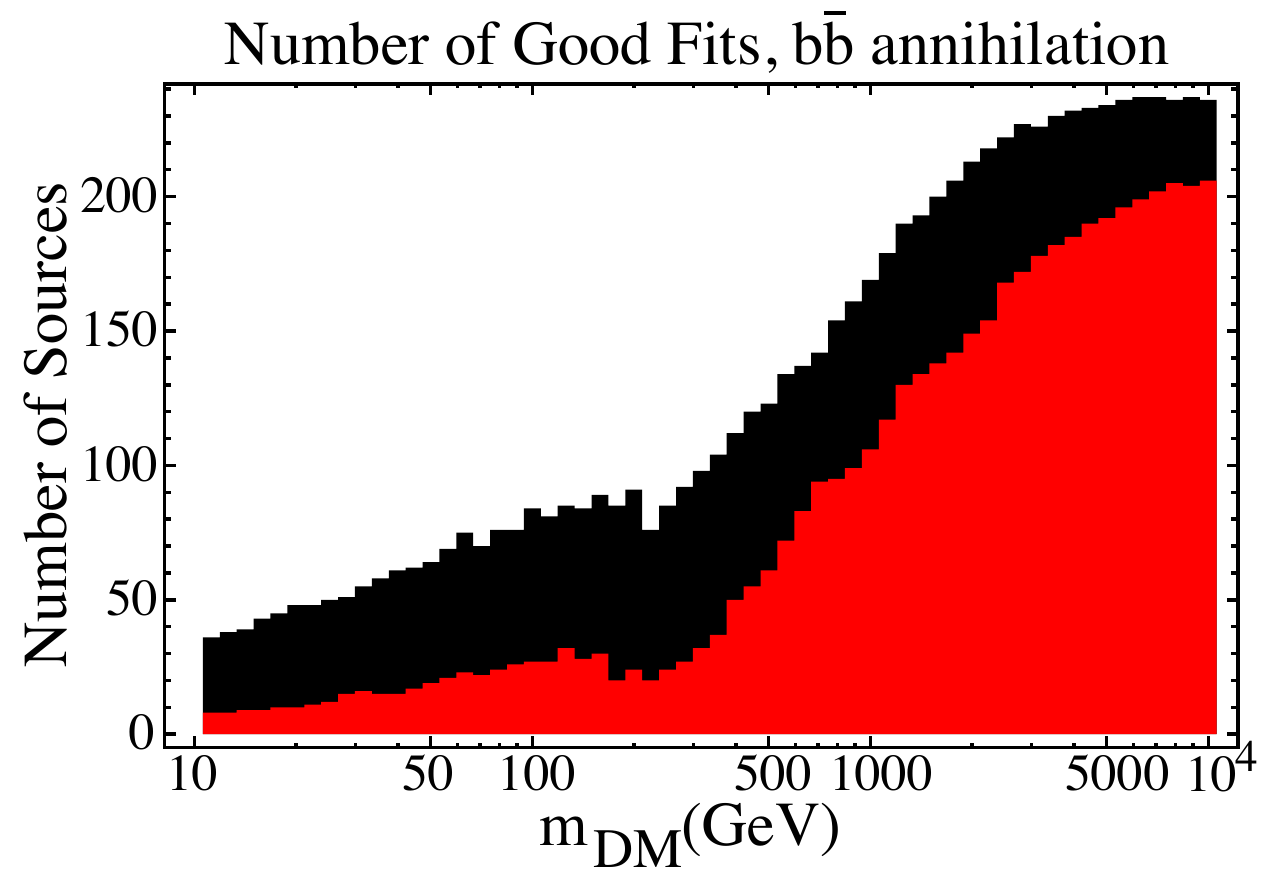}}
\hspace{0.02\linewidth}
{\includegraphics[angle=0,width=0.46\linewidth]{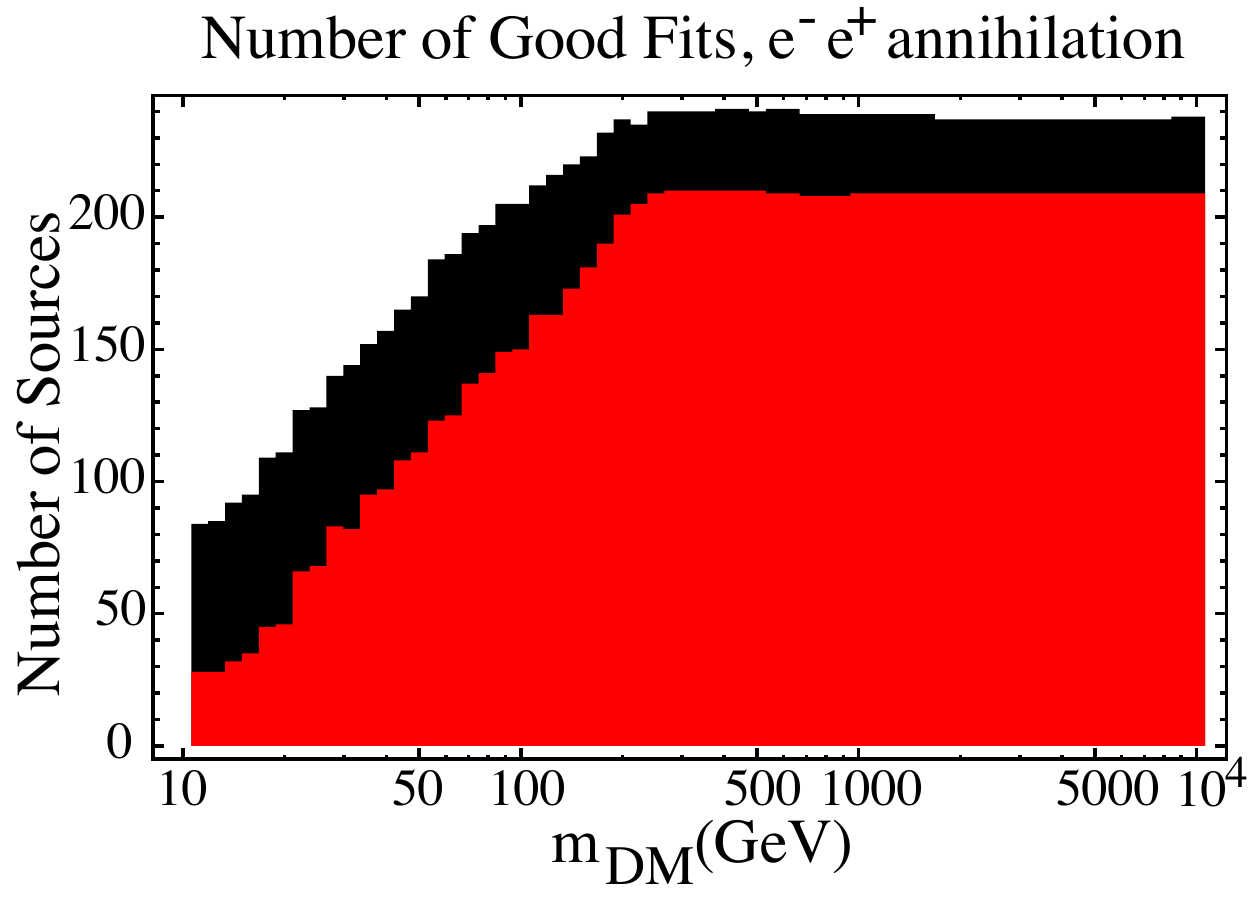}}\\
{\includegraphics[angle=0,width=0.46\linewidth]{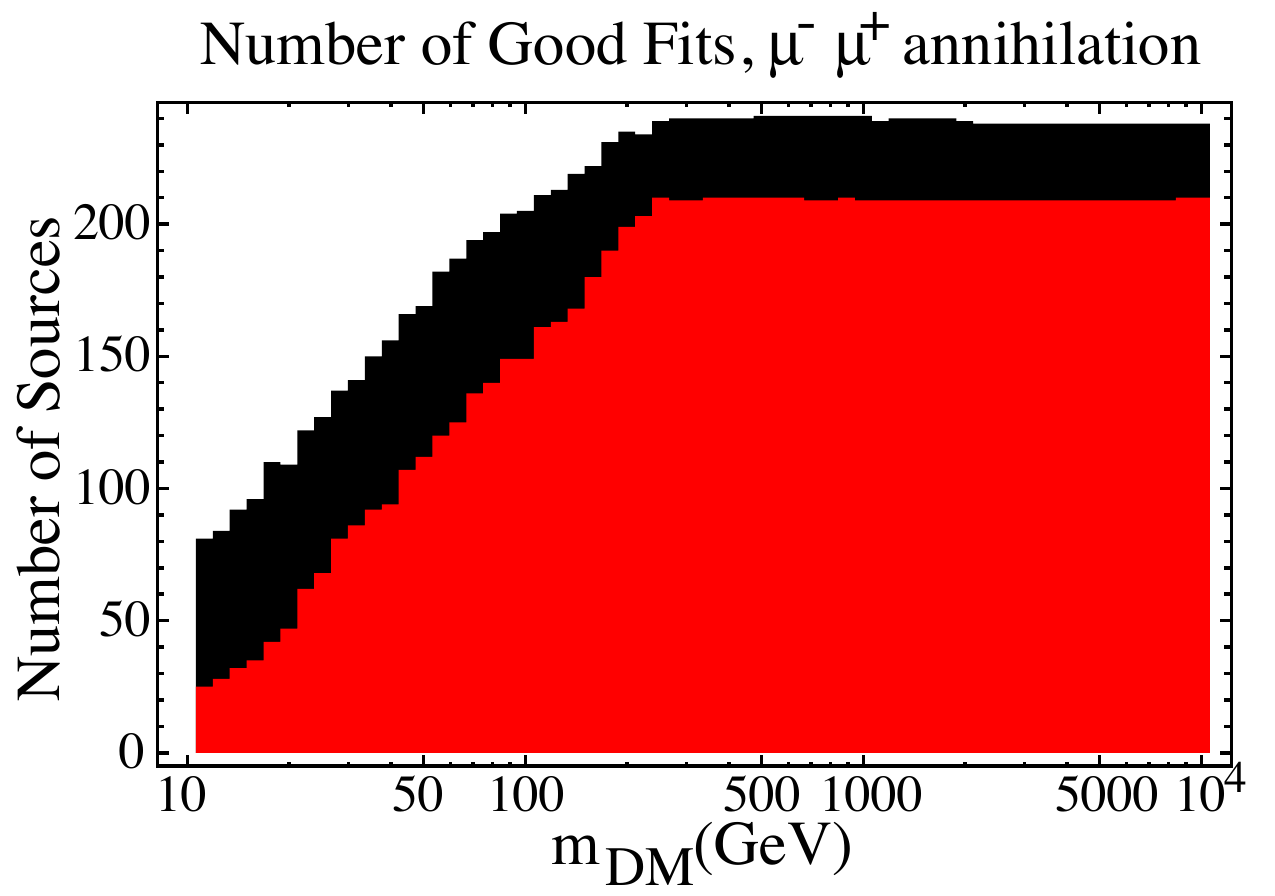}}
\hspace{0.02\linewidth}
{\includegraphics[angle=0,width=0.46\linewidth]{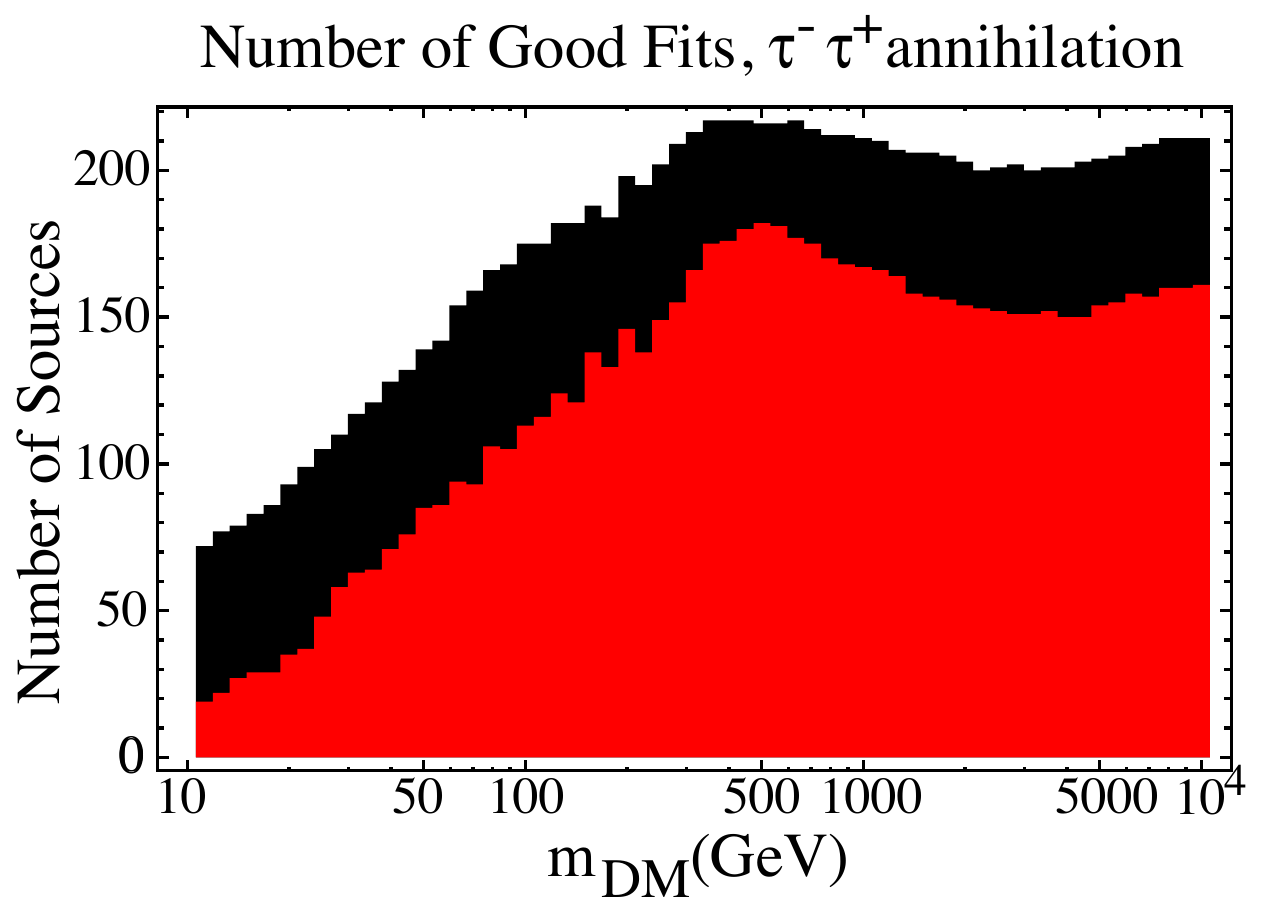}}\\
{\includegraphics[angle=0,width=0.46\linewidth]{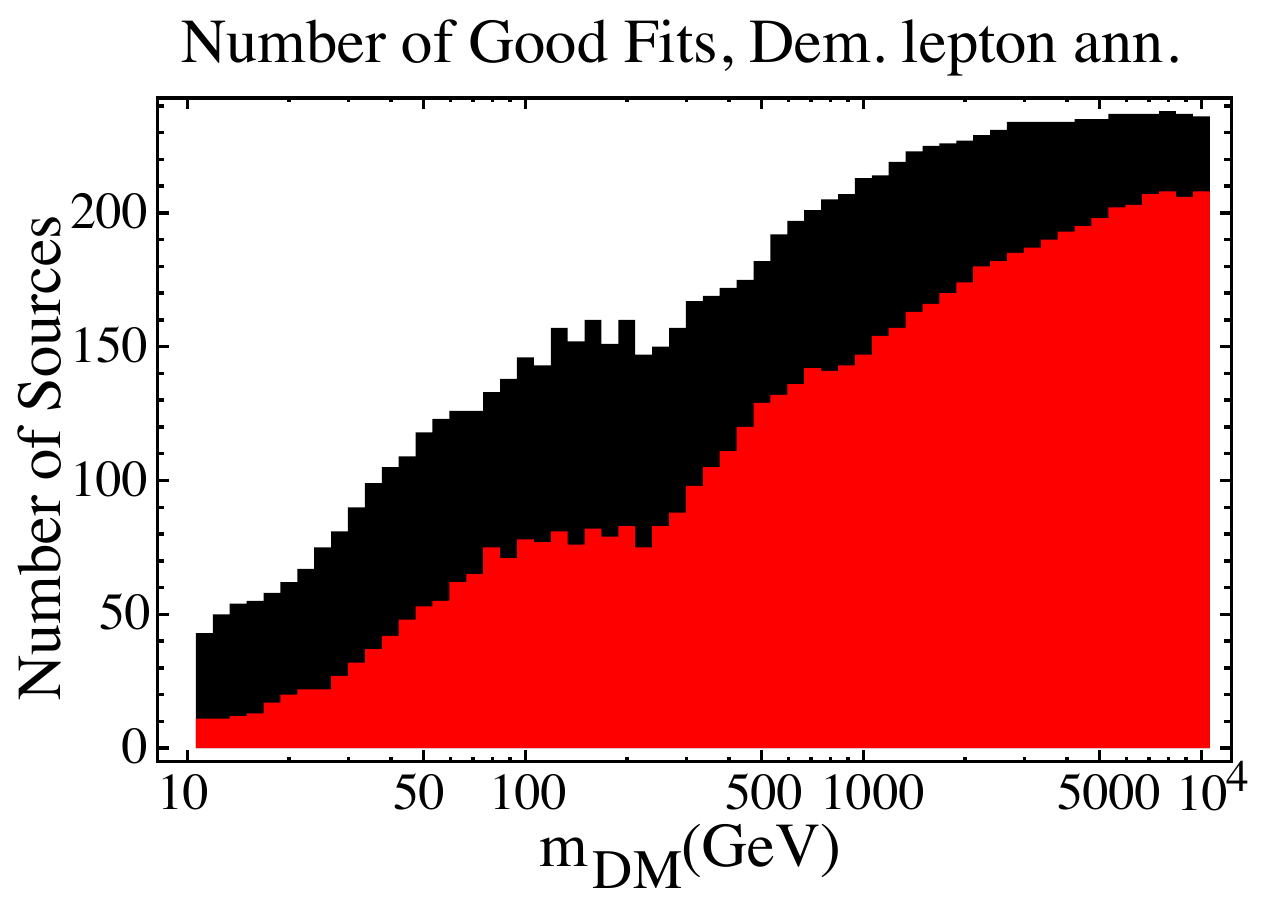}}
\hspace{0.02\linewidth}
{\includegraphics[angle=0,width=0.46\linewidth]{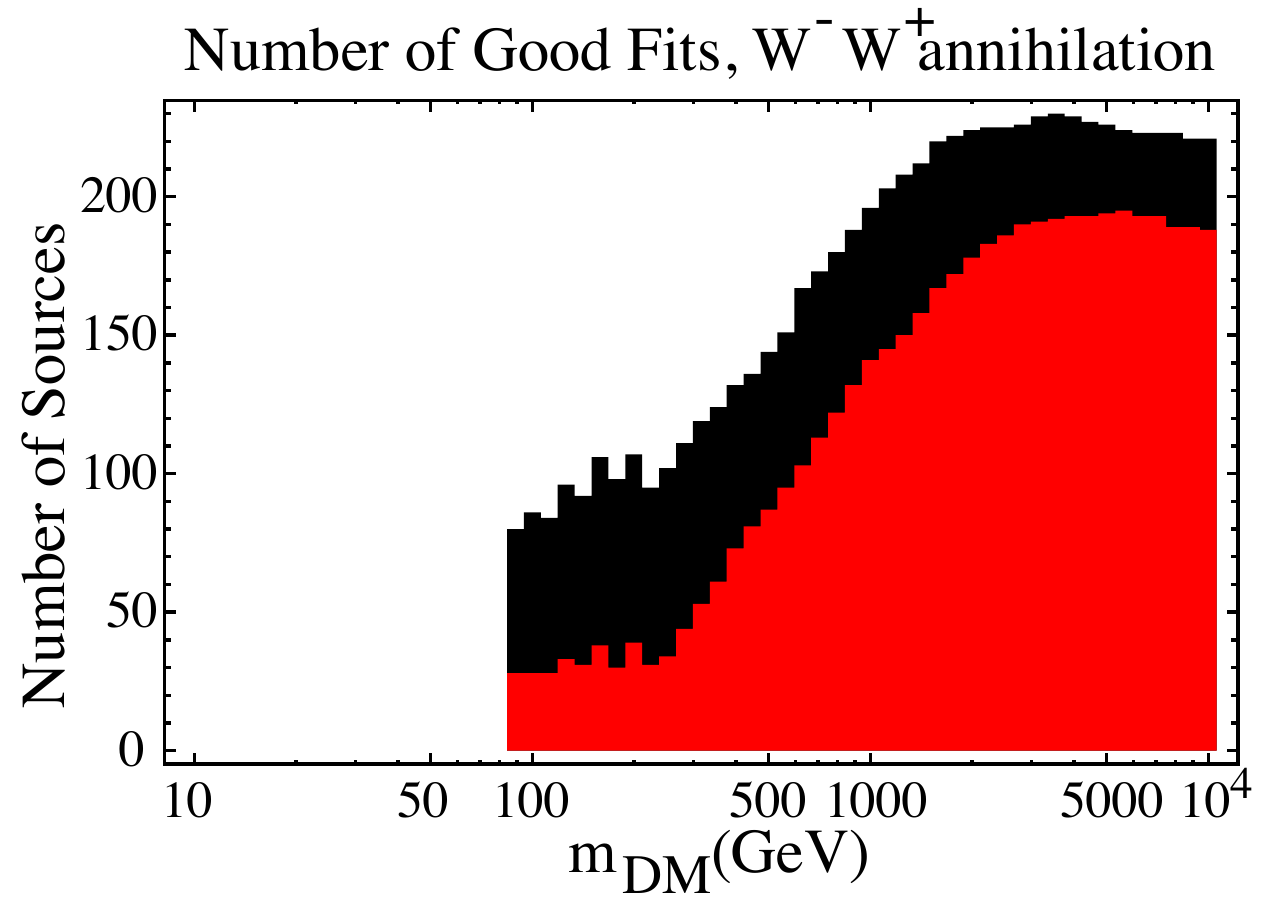}}
\caption{The number of dark matter subhalo candidate point sources in the Fermi First Source Catalog that can be well fit by annihilating dark matter with a given mass and dominant annihilation channel.  The red and black histograms denote fits with $\chi^2/$D.o.F.$<1.0$ and $<1.2$, respectively.}
\label{fig:goodfit}
\end{center}
\end{figure*}

Not surprisingly, we find that many of the 368 subhalo candidates in the Fermi First Source Catalog can be well fit by a spectrum from annihilating dark matter. In Fig.~\ref{fig:spectrumfits}, we show a few examples of such fits. Although the dark matter fits for each of these sources provides a good overall $\chi^2$, one can certainly not conclude from this information that any of these sources are a dark matter subhalo. If, on the other hand, it could be shown that a large number of subhalo candidate sources in the catalog could be well fit by a specific dark matter candidate (meaning a specific mass and combination of annihilation channels), then perhaps a case for a dark matter interpretation could be made. 

With this goal in mind, we plot in Fig.~\ref{fig:goodfit} the number of sources in the catalog that are well fit, meaning $\chi^2/$D.o.F.$<1$ (red) or 1.2 (black), by annihilating dark matter of a given mass and annihilation channel. Note that each point source can (and indeed does) appear many times in each histogram.

Examining these histograms, we notice a number of interesting features. Firstly, a large number of good fits are found for large dark matter masses. This is simply the result of few very high energy photons in the observed spectra of the sources. A very heavy dark matter particle, regardless of annihilation channel, can provide the required handful of $\sim 100-200$~GeV gamma rays, while negligibly contributing to the low energy flux. A class of astrophysical gamma ray sources with a very hard spectrum, however, could also fit the observed spectra of these sources. 

We also notice several bump-like features in the distributions shown in Fig.~\ref{fig:goodfit}. In particular, we observe features near $100-200$~GeV for the $b\bar{b}$, $W^+W^-$, and democratic lepton channels, and near $500$~GeV for the $\tau^+\tau^-$ channel. We will return to possible interpretations of these features in Sec.~\ref{hints}.

\section{Constraining The Dark Matter Annihilation Cross Section}
\label{limits}

In this section, we use the results shown in Fig.~\ref{fig:goodfit} to determine the maximum number of dark matter subhalos that could plausibly be present within the Fermi First Source Catalog, and use that information to calculate an upper limit on the dark matter's annihilation cross section.

\begin{figure*}[t]
\begin{center}
{\includegraphics[angle=0,width=0.48\linewidth]{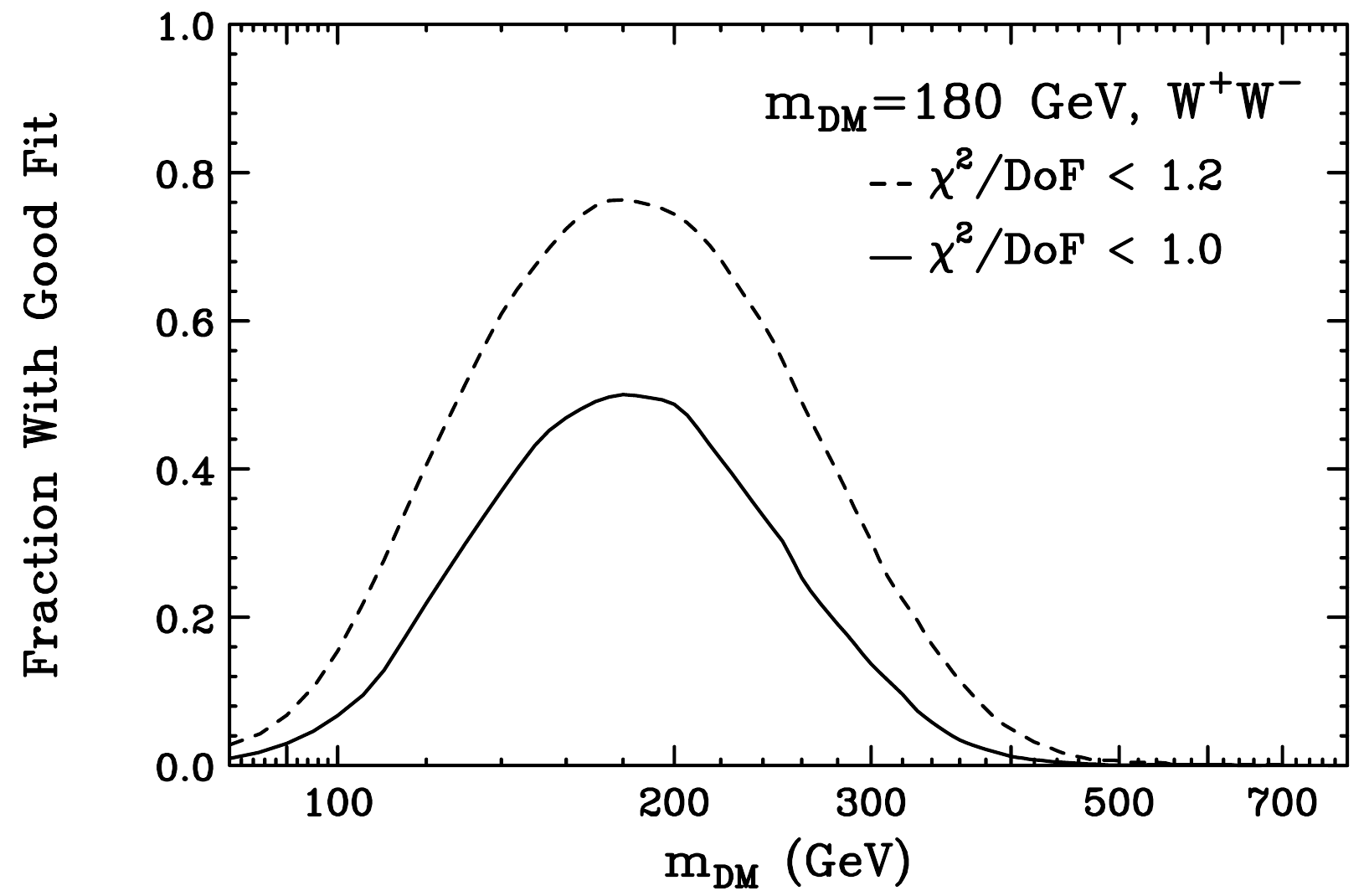}}
\hspace{0.02\linewidth}
{\includegraphics[angle=0,width=0.48\linewidth]{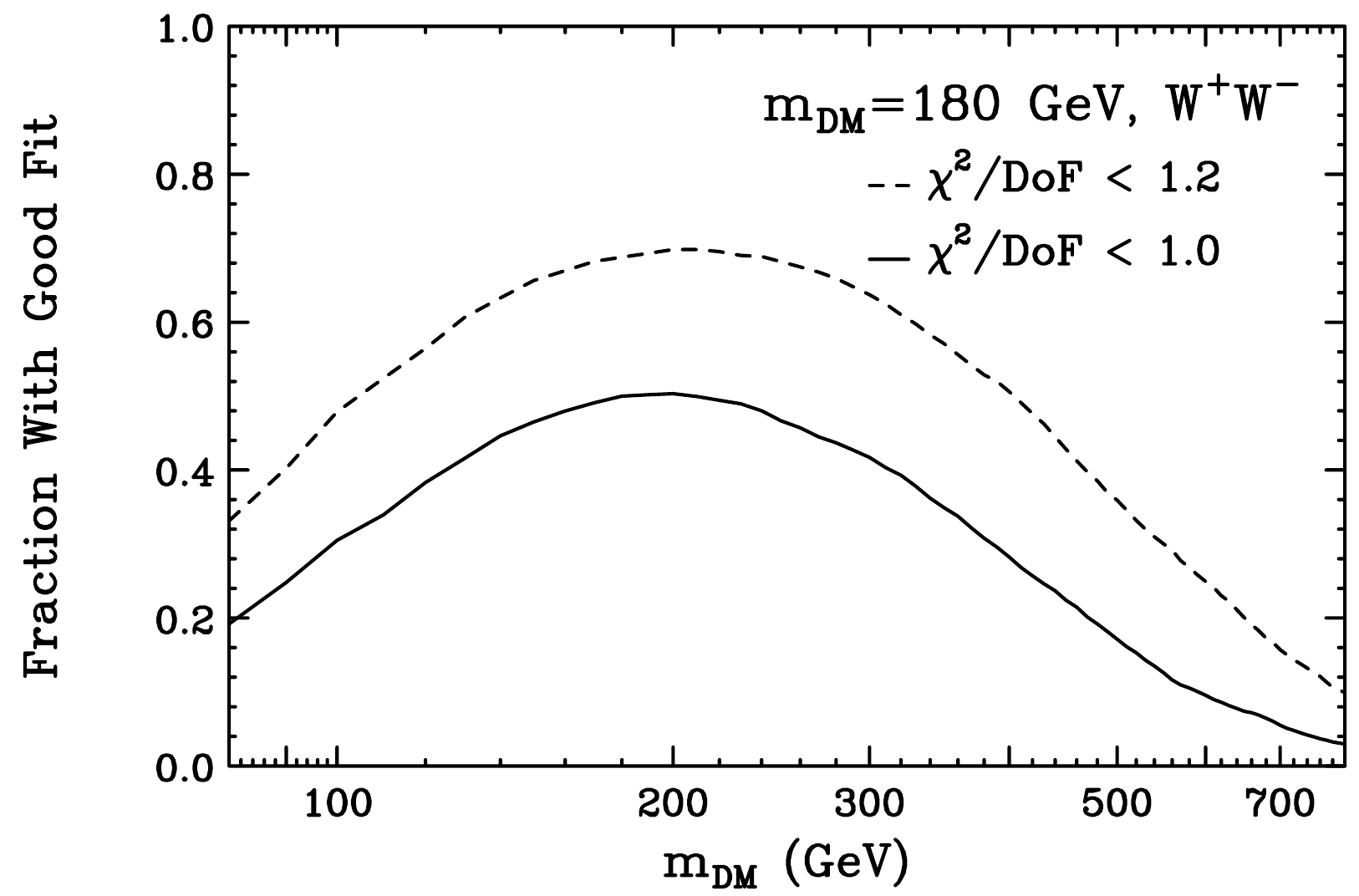}}
\vspace{-0.3cm}
\caption{The fraction of simulated subhalos (for the case example of $m_{\rm DM}$ annihilating to $W^+ W^-$, each producing 50 events above 1 GeV) that provide a good fit to a dark matter hypothesis, as a function of the dark matter's mass. In the left frame, no backgrounds are included. In the right frame, we have include a background of the form $dN_{\gamma}/dE_{\gamma} \propto E_{\gamma}^{-2}$ below 2 GeV, and $dN_{\gamma}/dE_{\gamma} \propto E_{\gamma}^{-2.6}$ above 2 GeV, normalized such that it produces 40 events per subhalo above 1 GeV (appropriate for diffuse galactic emission at latitudes of $20^{\circ}<|b|<60^{\circ}$).}
\label{MC}
\end{center}
\end{figure*}

To begin, we used a Monte Carlo to estimate what a signal from dark matter would potentially look like within the histograms of Fig.~\ref{fig:goodfit}. In Fig.~\ref{MC} we show, for the case of a 180 GeV dark matter particle that annihilates to $W^+ W^-$, the fraction of subhalos that are predicted to be fit well by an annihilating dark matter hypothesis, as a function of the dark matter's mass. Here we have considered point-like subhalos that produce 50 events above 1 GeV. As expected, the fraction with a good fit peaks at $m_{\rm DM}=180$ GeV, but with considerable width to the distribution. In the left frame, backgrounds were neglected, while in the right frame we adopted a diffuse gamma ray background of the form $dN_{\gamma}/dE_{\gamma} \propto E_{\gamma}^{-2}$ below 2 GeV, and $dN_{\gamma}/dE_{\gamma} \propto E_{\gamma}^{-2.6}$ above 2 GeV, normalized such that it produces 40 events above 1 GeV per subhalo. The inclusion of this background broadens the range of masses that can fit the simulated data significantly. At high galactic latitudes ($|b|>60^{\circ}$), where the galactic diffuse emission is lowest ($\sim$20 events per year per square degree above 1 GeV ), this distribution resembles that shown in the left frame. Over most of the sky ($20^{\circ} <|b|<60^{\circ}$), however, it will more closely resemble that shown in the right frame. For sources closer to the Galactic Plane, the distribution can be even broader. Note that brighter subhalos will also provide a distribution that is more strongly peaked at the dark matter's mass, although we expect most of the observable subhalos to be not much brighter than the threshold for their detection.

\begin{figure*}[t]
\begin{center}
{\includegraphics[angle=0,width=0.6\linewidth]{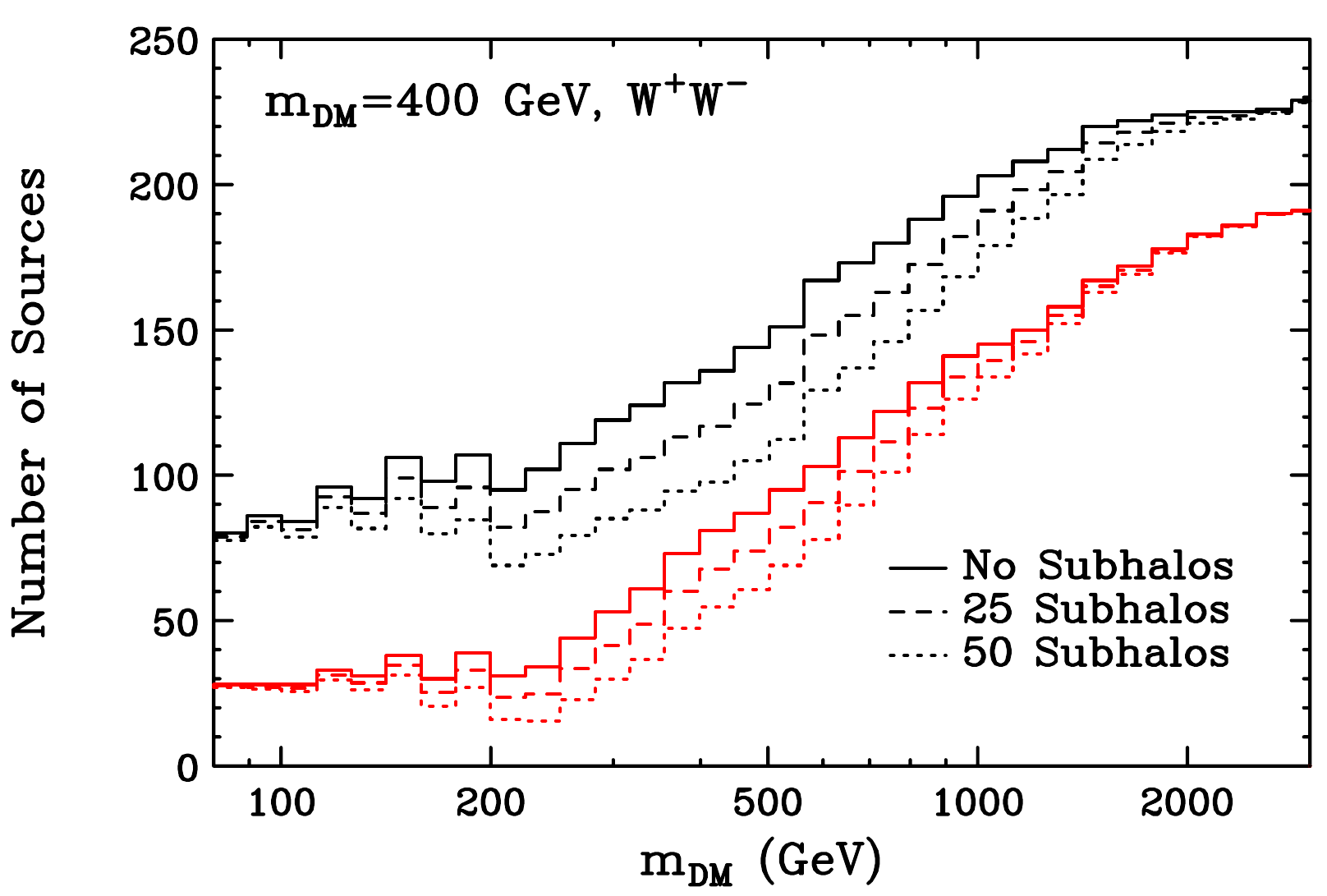}}
\vspace{-0.3cm}
\caption{The number of dark matter subhalo candidate point sources in the Fermi First Source Catalog that can be well fit by a dark matter particle annihilating to $W^+W^-$, as a function of the dark matter particle's mass (solid). Also shown is the result if a population of 25 (dashed) or 50 (dotted) subhalos is subtracted from the distribution, assuming a 400 GeV dark matter mass. When subtracting approximately 35 or more subhalos from the distribution, a depression-like feature begins to appear in the distribution, indication that an oversubtraction is taking place.}
\label{400}
\end{center}
\end{figure*}

Using these distributions, we then proceeded to calculate how many subhalos could possibly be contained within the histograms shown in Fig.~\ref{fig:goodfit}. At a minimum, we could simply conclude that there there are no more subhalos in the sample than provide good fits. For example, if we consider a 400 GeV dark matter particle annihilating to $W^+ W^-$, we see from Fig.~\ref{fig:goodfit} that about 72 (130) subhalo candidate sources in the Fermi First Source Catalog that provide a fit better than $\chi^2/$DoF $< 1.0$ (1.2). From this, we can robustly conclude that fewer than approximately 145 subhalos are present in the catalog. 

However, if we also consider the shapes of the distributions in Fig.~\ref{fig:goodfit}, we can in some cases more stringently constrain the number of subhalos contained in the catalog. In particular, if a very large number of subhalos were actually present within one of the histograms shown in Fig.~\ref{fig:goodfit}, then subtracting that population from the distribution observed would leave a histogram with a deep, depression-like feature. While it is possible that the dark matter signal distribution (such as that shown in Fig.~\ref{MC}) is exactly balanced by a depression in the distribution from astrophysical sources alone, we consider this unlikley and fine tuned. 


In Fig.~\ref{400}, we demonstrate how we can use this approach to more stringently constrain the number of subhalos in the Fermi First Source Catalog. Here, we plot the histogram for the $W^+W^-$ channel, subtracting from it a population of 0, 25, or 50 subhalos (consisting of dark matter particles with $m_{\rm DM}=400$ GeV). By requiring that the subtraction of a subhalo population does not lead to a statistically significant depression-like feature in the histograms, we can often place a more stringent limit the number of subhalos present in the Fermi First Source Catalog. In the case of a 400 GeV dark matter particle, annihilating to $W^+ W^-$, for example, we conclude that less than approximately 35 subhalos are present in the catalog.

\begin{figure*}[t]
\begin{center}
{\includegraphics[angle=0,width=0.8\linewidth]{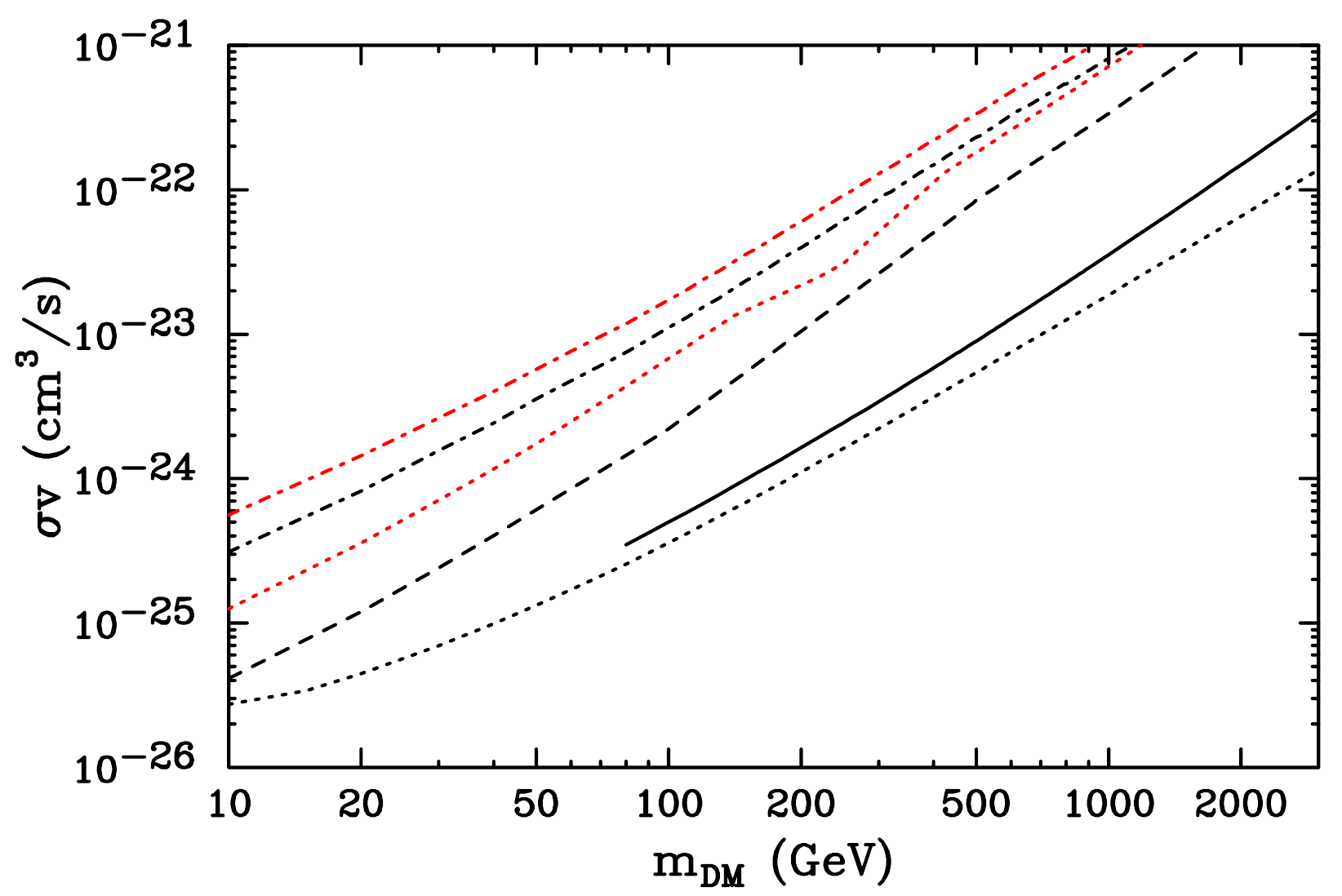}}
\vspace{-0.3cm}
\caption{The 95\% confidence level upper limits on the dark matter's annihilation cross section from studies of the Fermi First Source Catalog. From top-to-bottom, the curves denote dark matter annihilating to $\mu^+ \mu^-$ (dot-dashed red), $e^+ e^-$ (dot-dashed black), democratic leptons (dotted red), $\tau^+ \tau^-$ (dashed black), $W^+ W^-$ (solid black), and $b\bar{b}$ (dotted black). We have used the default assumptions as discussed in Sec.~\ref{theory} ($\gamma=1.2$, 99\% mass loss, no sub-subhalos). See text for more details.}
\label{limit}
\end{center}
\end{figure*}

We have repeated this procedure for each dark matter mass and annihilation channel and determined in each case the maximum number of dark matter subhalos possibly contained within the Fermi First Source Catalog. From the maximum number of subhalos, we then calculated an upper limit on the dark matter's annihilation cross section. In Fig.~\ref{limit}, we show the upper limit on the annihilation cross section as a function of mass, and for various annihilation channels. Here we have used our default assumptions as discussed in Sec.~\ref{theory} ($\gamma=1.2$, 99\% mass loss, no sub-subhalos), and assumed that the Fermi First Source Catalog is approximately complete for sources that produce more than $\sim$50 events per year above 1 GeV. Although effects such as confusion between multiple point sources are expected to cause some sources to not appear in the FGST point source catalog, this is thought to impact less than 10\% of sources with $|b| > 10^{\circ}$~\cite{catalog}.

Of course, if we adopt different assumptions pertaining to the dark matter subhalo profiles, concentrations, and mass losses, we can arrive at somewhat different conclusions. Erring on the optimistic side, if we use our default profile shape, but with only 90\% mass loss, and include the effects of halo-to-halo variations in subhalo concentrations, our predicted number of observable subhalos increases by a factor of $\sim$5--6, corresponding to a limit on the annihilation cross section that is more stringent by a factor of approximately 3 (or even more stringent if subhalos have their own substructure). On the other hand, if we conservatively adopt a less cusped profile shape ($\gamma=1.0$), 99\% mass loss, and neglect halo-to-halo variations, we arrive at a limit that is about 4 times less stringent than that shown in Fig.~\ref{limit}.

Comparing these results to other constraints that have been placed on the dark matter annihilation cross section, we find that our constraint is comparable to, or in some cases more stringent than, those found previously. In particular, the FGST collaboration has published constraints from observations of 14 dwarf spheroidal galaxies (Draco and Ursa Minor providing the most stringent constraints)~\cite{fgstdwarf}. For a 30 GeV dark matter particle annihilating to $b\bar{b}$, for example, that study concludes that $\sigma v < 7 \times 10^{-26}$ cm$^3$/s, which is comparable to our result shown in Fig.~\ref{limit}. For heavier dark matter masses, however, the limits from dwarf spheroidals are somewhat more stringent (by a factor of $\sim 5$ at 1 TeV, for example). For masses below a few hundred GeV, our results also provide a more stringent limit than those from observations of galactic and extragalactic diffuse emission~\cite{Abdo:2010dk}.

\section{Possible Hints Of Dark Matter In The Fermi First Source Catalog?}
\label{hints}

\begin{figure*}[t]
\begin{center}
{\includegraphics[angle=0,width=0.48\linewidth]{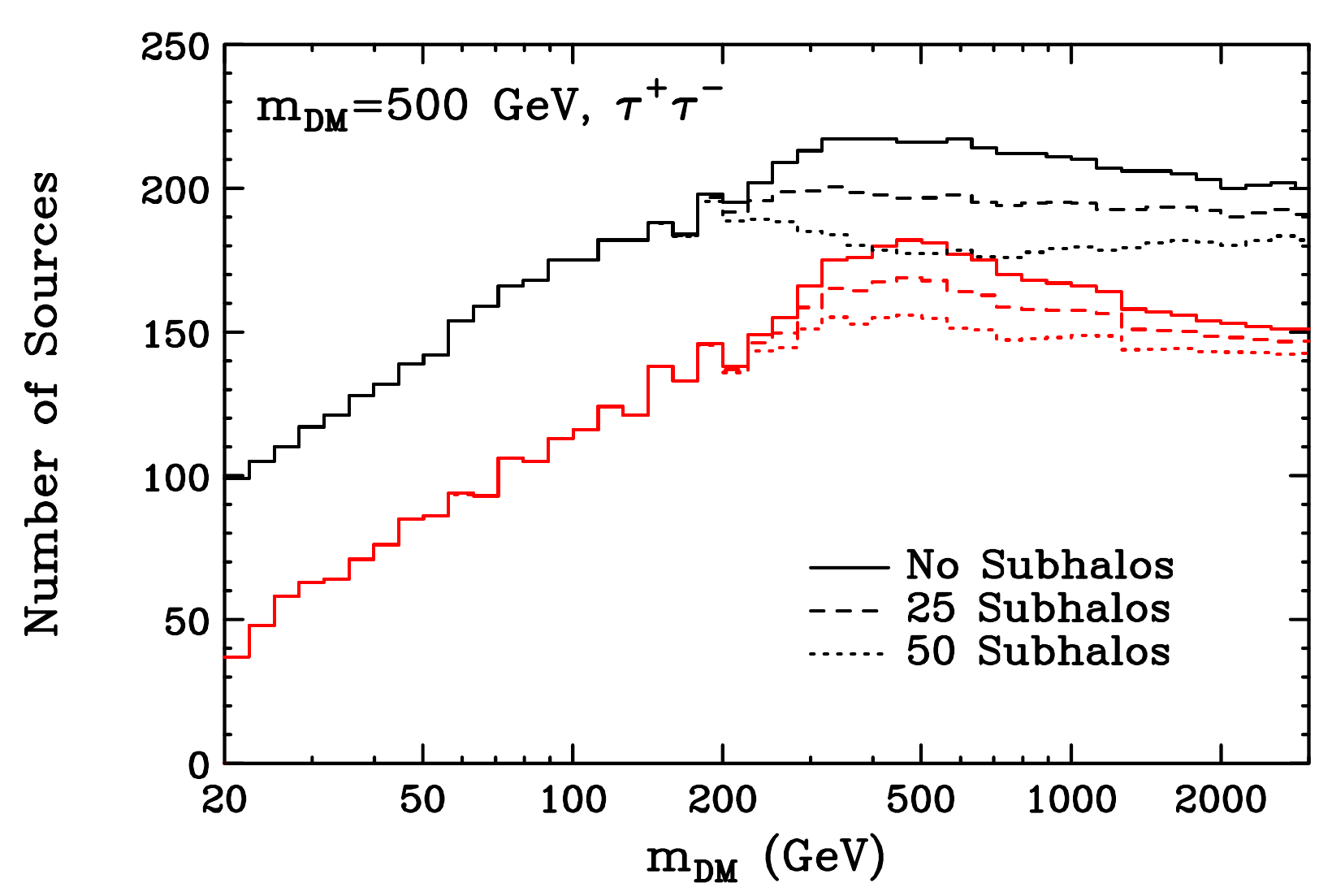}}
\hspace{0.02\linewidth}
{\includegraphics[angle=0,width=0.48\linewidth]{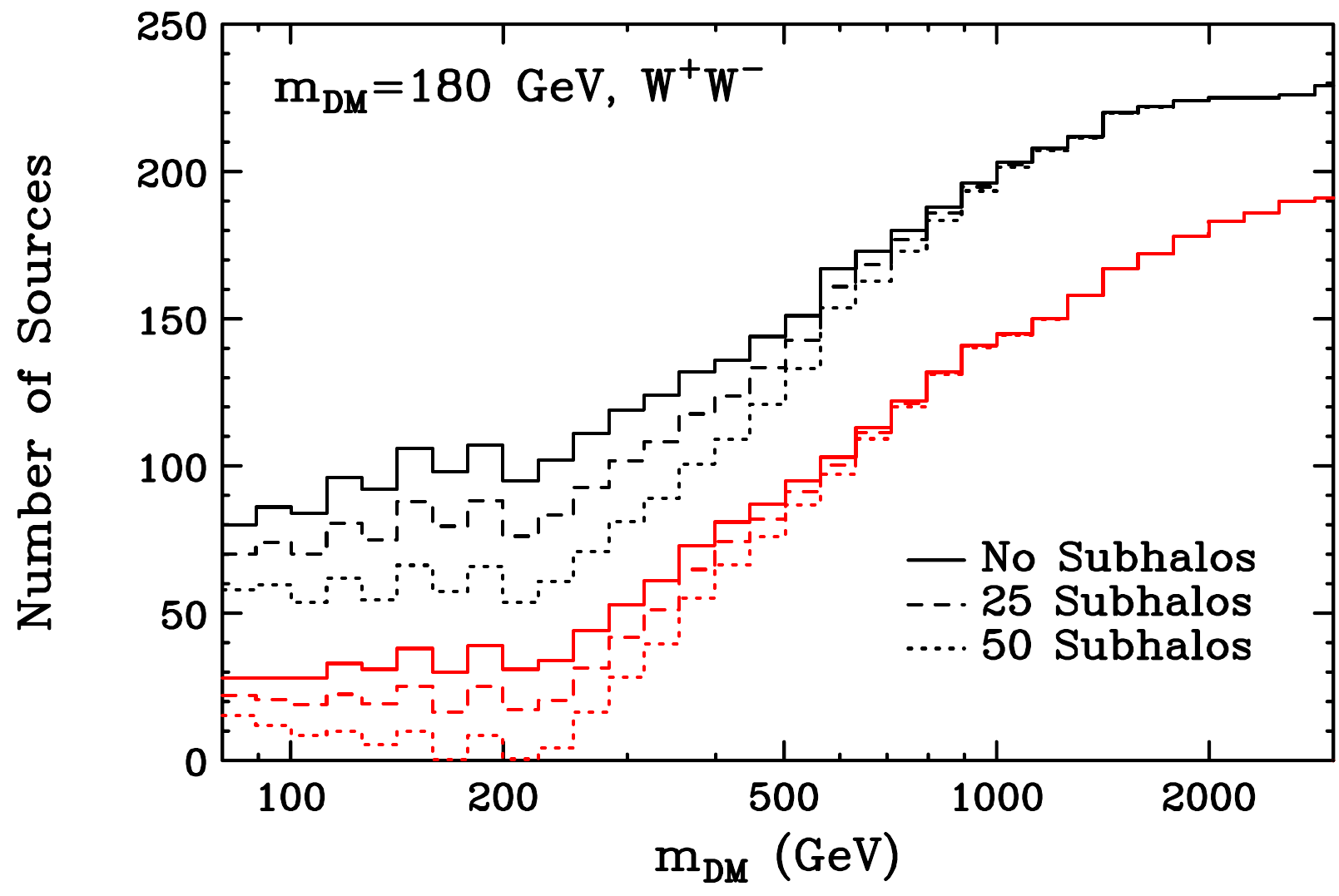}}
\vspace{-0.3cm}
\caption{The number of dark matter subhalo candidate point sources in the Fermi First Source Catalog that can be well fit by a dark matter particle annihilating to $\tau^+ \tau^-$ (left) or $W^+W^-$ (right), as a function of the dark matter particle's mass (solid). Also shown in each frame is the result if a population of 25 (dashed) or 50 (dotted) subhalos is subtracted from the distribution, assuming a 500 GeV or 180 GeV dark matter mass in the left and right frames, respectively. When subtracting approximately 30 subhalos from either distribution, the observed bump-like features are flattened. The observed bump-like feature could thus be explained if approximately 30 of the sources within the Fermi First Source Catalog are dark matter subhalos.}
\label{subtract}
\end{center}
\end{figure*}

In this section, we turn our attention to some of the features observed in the histograms of Fig.~\ref{fig:goodfit}, and explore the possibility that perhaps $\sim$10--50 of the sources in the Fermi First Source Catalog might be dark matter subhalos. 

In particular, we notice in Fig.~\ref{fig:goodfit} potentially interesting bump-like features at $m_{\rm DM}\sim 500$ GeV, for annihilations to $\tau^+ \tau^-$, and at $m_{\rm DM}\sim 150-200$ GeV, for annihilations to $W^+ W^-$, $b \bar{b}$, or democratic leptons. Could these features be the result of a population of dark matter subhalos within the FGST point source catalog? To assess this question, we subtracted from the observed distribution of sources that which would be predicted (by our Monte Carlo) from a dark matter subhalo population. In Fig.~\ref{subtract}, we show example results of this subtraction. For the case of a 500 GeV dark matter particle annihilating to $\tau^+ \tau^-$, we find that the observed bump can be removed, flattening the overall distribution, if a population of approximately 30 subhalos is present within the catalog. A similar number of subhalos with $m_{\rm DM}=180$ GeV, annihilating to either $W^+ W^-$ or $b\bar{b}$, would also also largely removed the features seen in their respective channels. 

In the $m_{\rm DM}=500$ GeV, $\tau^+ \tau^-$ case, in order to produce the required number of observable subhalos, the dark matter must have a very large annihilation cross section ($\sigma v \sim 6 \times 10^{-23}$ cm$^3$/s). Intriguingly, in order for such a dark matter particle to provide the excess positron fraction reported by PAMELA~\cite{pamela}, it must have a cross section of approximately $\sigma v \sim 2 \times 10^{-23}$~\cite{boostpam}. The factor of $\sim$3 disparity between these two values could easily be accounted for if, for example, subhalos contain significant substructure, or less mass loss has taken place than we had assumed (see Table~\ref{table}).

In the case of $m_{\rm DM}=180$ GeV, and annihilations to $W^+ W^-$, we find again that a population of approximately 30 subhalos within the Fermi First Source Catalog can flatten the observed distribution. This corresponds to an annihilation cross section of $\sim (1-2) \times 10^{-24}$ cm$^3$/s. This is comparable to the value predicted for a wino-like neutralino, such as is found in models of anomaly mediated supersymmetry breaking~\cite{wino}. A similar cross section would also be found for the case of annihilations to $b\bar{b}$. We do not consider the democratic lepton case further as this appears to conflict with the cosmic ray electron spectrum as measured by FGST~\cite{inter}.

We emphasize that none these features can, at this time, be used to confidently support a claim that a population of dark matter subhalos is present within the Fermi First Source Catalog. It is the case, however, that the presence of such a population could explain some of the features observed in the catalog.  If either of these observed features does, in fact, result from the presence of a subhalo population, the feature is be expected to become more prominent as the FGST accumulates more data, and the subhalos' spectra become more precisely measured.

\section{Summary and Conclusions}
\label{conclusions}

In this article, we have studied the recently published Fermi Gamma Ray Space Telescope's First Source Catalog, and considered whether a significant number of the sources in this catalog might be dark matter subhalos. For a typical thermal dark matter candidate with a mass of 50 GeV, and reasonable astrophysical assumptions, one predicts that a few relatively large ($\sim 10^3-10^7 M_{\odot}$) and nearby ($\sim 0.01-10$ kpc) subhalos would be bright enough to appear within such a catalog. It is unlikely, however, that such a small number of sources could be identified from among the hundreds of unidentified sources in the Fermi First Source Catalog. If the dark matter's annihilation cross section were larger than that typically predicted for a thermal relic, however, it is possible that a larger number of subhalos could be detectable by gamma ray telescopes. 

From among Fermi's First Source Catalog, we studied 368 sources as dark matter subhalo candidates, each of which has been detected with high significance ($> 5 \sigma$), is located away from the Galactic Plane ($|b| > 10^{\circ}$), and is not associated with any source observed at other wavelengths. Although we find that the spectrum of many of these sources could be well fit by that predicted from annihilating dark matter, this is most likely the result of the relatively large error bars involved, and on the wide range of dark matter masses and annihilation channels we have considered. 

By studying the distribution of source spectra in the First Source Catalog, we derived an upper limit on the annihilation cross section of dark matter, as a function of its mass and dominant annihilation channel. For dark matter particles with relatively low masses (less than a few hundred GeV), we find constraints that are comparably stringent to those derived from observations of dwarf spheroidal galaxies. For example, in the case of a 50 GeV dark matter particle that annihilates to $b\bar{b}$, we conclude that $\sigma v \lsim 7 \times 10^{-26}$ cm$^3$/s. 

Lastly, we note the appearance of bump-like features in the distribution of spectra observed in the Fermi First Source Catalog. Such features could be explained ({\it ie.} flattened) if the dark matter takes the form of a 500 GeV particle with an annihilation cross section to $\tau^+ \tau^-$ of $\sigma v \sim 6 \times 10^{-23}$ cm$^3$/s, or a 180 GeV particle with an annihilation cross section to $W^+ W^-$ of $\sigma v \sim (1-2) \times 10^{-24}$ cm$^2$/s (such as is predicted for a wino-like neutralino). These scenarios could also potentially account for the cosmic positron excess observed by PAMELA.

\bigskip

{\it Acknowledgements:} MB is supported by the US Department of Energy, under grant DE-FG03-92-ER40701. DH is supported by the US Department of Energy, including grant DE-FG02-95ER40896, and by NASA grant NAG5-10842.


\begin{thebibliography}{99}


\bibitem{Pieri:2007ir}
  L.~Pieri, G.~Bertone and E.~Branchini,
  Mon.\ Not.\ Roy.\ Astron.\ Soc.\  {\bf 384}, 1627 (2008)
  [arXiv:0706.2101 [astro-ph]].

\bibitem{catalog}
  T.~L.~Collaboration,
  arXiv:1002.2280 [astro-ph.HE].


\bibitem{norm}
  J.~Diemand, M.~Kuhlen and P.~Madau,
  Astrophys.\ J.\  {\bf 657}, 262 (2007)
  [arXiv:astro-ph/0611370];
 M.~Fornasa, L.~Pieri, G.~Bertone and E.~Branchini,
  Phys.\ Rev.\  D {\bf 80}, 023518 (2009)
  [arXiv:0901.2921 [astro-ph]].

\bibitem{smallest}
  S.~Profumo, K.~Sigurdson and M.~Kamionkowski,
  Phys.\ Rev.\ Lett.\  {\bf 97}, 031301 (2006)
  [arXiv:astro-ph/0603373];
  A.~Loeb and M.~Zaldarriaga,
  Phys.\ Rev.\  D {\bf 71}, 103520 (2005)
  [arXiv:astro-ph/0504112].


\bibitem{1pt2}
  J.~Diemand, M.~Kuhlen, P.~Madau, M.~Zemp, B.~Moore, D.~Potter and J.~Stadel,
  Nature {\bf 454}, 735 (2008)
  [arXiv:0805.1244 [astro-ph]];
  J.~Diemand, M.~Zemp, B.~Moore, J.~Stadel and M.~Carollo,
  Mon.\ Not.\ Roy.\ Astron.\ Soc.\  {\bf 364}, 665 (2005)
  [arXiv:astro-ph/0504215].

\bibitem{shallow}
 J.~F.~Navarro {\it et al.},
  Mon.\ Not.\ Roy.\ Astron.\ Soc.\  {\bf 349}, 1039 (2004)
  [arXiv:astro-ph/0311231];
  V.~Springel {\it et al.},
  Mon.\ Not.\ Roy.\ Astron.\ Soc.\  {\bf 391}, 1685 (2008)
  [arXiv:0809.0898 [astro-ph]].


\bibitem{einasto}
  D.~Merritt, J.~F.~Navarro, A.~Ludlow and A.~Jenkins,
  Astrophys.\ J.\  {\bf 624}, L85 (2005)
  [arXiv:astro-ph/0502515].

\bibitem{bullock}
  J.~S.~Bullock {\it et al.},
  Mon.\ Not.\ Roy.\ Astron.\ Soc.\  {\bf 321}, 559 (2001)
  [arXiv:astro-ph/9908159].

\bibitem{Kuhlen:2008aw}
  M.~Kuhlen, J.~Diemand and P.~Madau,
  arXiv:0805.4416 [astro-ph].


\bibitem{disruption}
  H.~Zhao, J.~E.~Taylor, J.~Silk and D.~Hooper,
  Astrophys.\ J.\  {\bf 654}, 697 (2007)
  [arXiv:astro-ph/0508215];
  arXiv:astro-ph/0502049;
  T.~Goerdt, O.~Y.~Gnedin, B.~Moore, J.~Diemand and J.~Stadel,
  Mon.\ Not.\ Roy.\ Astron.\ Soc.\  {\bf 375}, 191 (2007)
  [arXiv:astro-ph/0608495];
  A.~M.~Green and S.~P.~Goodwin,
  Mon.\ Not.\ Roy.\ Astron.\ Soc.\  {\bf 375}, 1111 (2007)
  [arXiv:astro-ph/0604142];
  V.~Berezinsky, V.~Dokuchaev and Y.~Eroshenko,
  Phys.\ Rev.\  D {\bf 73}, 063504 (2006)
  [arXiv:astro-ph/0511494].

\bibitem{latper}
\url{http://www-glast.slac.stanford.edu/software/IS/glast_lat_performance.htm}

\bibitem{pythia}
  T.~Sjostrand, P.~Eden, C.~Friberg, L.~Lonnblad, G.~Miu, S.~Mrenna and E.~Norrbin,
  Comput.\ Phys.\ Commun.\  {\bf 135}, 238 (2001)
  [arXiv:hep-ph/0010017].
  
\bibitem{Gondolo:2004sc}
  P.~Gondolo, J.~Edsjo, P.~Ullio, L.~Bergstrom, M.~Schelke and E.~A.~Baltz,
  JCAP {\bf 0407}, 008 (2004)
  [arXiv:astro-ph/0406204].
  

\bibitem{Bergstrom:2004cy}
  L.~Bergstrom, T.~Bringmann, M.~Eriksson and M.~Gustafsson,
  Phys.\ Rev.\ Lett.\  {\bf 94}, 131301 (2005)
  [arXiv:astro-ph/0410359].


\bibitem{subsolar}
  S.~K.~Lee, S.~Ando and M.~Kamionkowski,
  JCAP {\bf 0907}, 007 (2009)
  [arXiv:0810.1284 [astro-ph]];
 S.~M.~Koushiappas,
  New J.\ Phys.\  {\bf 11}, 105012 (2009)
  [arXiv:0905.1998 [astro-ph.CO]].


\bibitem{fgstdwarf}
  A.~A.~Abdo {\it et al.},
  arXiv:1001.4531 [astro-ph.CO].


\bibitem{Abdo:2010dk}
  A.~A.~Abdo {\it et al.}  [Fermi-LAT Collaboration],
  arXiv:1002.4415 [astro-ph.CO];
  K.~N.~Abazajian, P.~Agrawal, Z.~Chacko and C.~Kilic,
  arXiv:1002.3820 [astro-ph.HE].



\bibitem{pamela}
  O.~Adriani {\it et al.}  [PAMELA Collaboration],
  Nature {\bf 458}, 607 (2009)
  [arXiv:0810.4995 [astro-ph]].


\bibitem{boostpam}
  P.~Meade, M.~Papucci, A.~Strumia and T.~Volansky,
  Nucl.\ Phys.\  B {\bf 831}, 178 (2010)
  [arXiv:0905.0480 [hep-ph]];
  D.~Hooper and K.~M.~Zurek,
  arXiv:0909.4163 [hep-ph];
  I.~Cholis, L.~Goodenough, D.~Hooper, M.~Simet and N.~Weiner,
  Phys.\ Rev.\  D {\bf 80}, 123511 (2009)
  [arXiv:0809.1683 [hep-ph]].

\bibitem{wino}
  T.~Moroi and L.~Randall,
  Nucl.\ Phys.\  B {\bf 570}, 455 (2000)
  [arXiv:hep-ph/9906527];
  G.~Kane, R.~Lu and S.~Watson,
  Phys.\ Lett.\  B {\bf 681}, 151 (2009)
  [arXiv:0906.4765 [astro-ph.HE]].

\bibitem{inter}
  D.~Grasso {\it et al.}  [FERMI-LAT Collaboration],
  Astropart.\ Phys.\  {\bf 32}, 140 (2009)
  [arXiv:0905.0636 [astro-ph.HE]].



\end{thebibliography}
\end{document}